\documentclass{elsart}
\usepackage{epsfig}
\usepackage{float}
\usepackage{lscape}
\journal{Nuc. Phys. A}

\begin{document}	


\begin{frontmatter}
\title{Sub-threshold $\phi$-meson yield in central
\nuc{58}{Ni}+\nuc{58}{Ni} collisions}

\author[firenze]{A.~Mangiarotti\thanksref{corauthor}},
\thanks[corauthor] {Corresponding author, address:
Largo E.Fermi 2, 50125 Firenze, Italy\\
Tel.: +39 055 2307787
FAX:  +39 055 229330
email: mangiaro@fi.infn.it}
\author[heidelberg]{N.~Herrmann},
\author[firenze]{P.~R.~Maurenzig},
\author[gsi]{A.~Gobbi},
\author[dresde]{R.~Kotte},
\author[budapest]{J.~Kecskemeti},
\author[gsi]{Y.~Leifels},
\author[clt]{J.~P.~Alard},
\author[gsi]{A.~Andronic},
\author[gsi]{R.~Averbeck},
\author[clt]{V.~Barret},
\author[zagreb]{Z.~Basrak},
\author[clt]{N.~Bastid},
\author[itep]{I.~Belyaev},
\author[clt]{A.~Bendarag},
\author[budapest]{G.~Berek},
\author[zagreb]{R.~\v{C}aplar},
\author[clt]{P.~Crochet},
\author[gsi]{A.~Devismes},
\author[clt]{P.~Dupieux},
\author[zagreb]{M.~D\v{z}elalija},
\author[gsi]{Ch.~Finck},
\author[budapest]{Z.~Fodor},
\author[itep]{Y.~Grishkin},
\author[gsi]{O.~Hartmann},
\author[gsi]{K.~D.~Hildenbrand},
\author[korea]{B.~Hong},
\author[korea]{Y.~J.~Kim},
\author[warsaw]{M.~Kirejczyk},
\author[gsi]{P.~Koczo{\'n}},
\author[zagreb]{M.~Korolija},
\author[gsi]{T.~Kress},
\author[gsi]{R.~Kutsche},
\author[itep]{A.~Lebedev},
\author[kur]{V.~Manko},
\author[heidelberg]{M.~Merschmeyer},
\author[bucarest]{D.~Moisa},
\author[kur]{A.~Nianine},
\author[dresde]{W.~Neubert},
\author[heidelberg]{D.~Pelte},
\author[bucarest]{M.~Petrovici},
\author[dresde]{C.~Plettner},
\author[ires]{F.~Rami},
\author[gsi]{W.~Reisdorf},
\author[ires]{B.~de~Schauenburg},
\author[gsi]{D.~Sch{\"u}ll},
\author[budapest]{Z.~Seres},
\author[warsaw]{B.~Sikora},
\author[korea]{K.~S.~Sim},
\author[bucarest]{V.~Simion},
\author[warsaw]{K.~Siwek-Wilczy{\'n}ska},
\author[itep]{V.~Smolyankin},
\author[heidelberg]{M.~Stockmeier},
\author[bucarest]{G.~Stoicea},
\author[kur]{M.~Vasiliev},
\author[ires]{P.~Wagner},
\author[heidelberg]{K.~Wi{\'s}niewski},
\author[dresde]{D.~Wohlfarth},
\author[kur]{I.~Yushmanov} and
\author[itep]{A.~Zhilin}

\address[bucarest]{National Institute for Nuclear Physics and
Engineering, Bucharest, Romania}
\address[budapest]{KFKI Research Institute for Particle and Nuclear
Physics, Budapest, Hungary}
\address[clt]{Laboratoire de Physique Corpusculaire, IN2P3/CNRS
and Universit\'e Blaise Pascal, Clermont-Ferrand, France}
\address[gsi]{Gesellschaft f\"ur Schwerionenforschung, Darmstadt,
Germany}
\address[dresde]{Forschungszentrum Rossendorf, Dresden, Germany}
\address[firenze]{Universit\`a di Firenze and Sezione INFN, Firenze,
Italy}
\address[heidelberg]{Physikalisches Institut der Universit\"at
Heidelberg, Heidelberg, Germany}
\address[itep]{Institute for Theoretical and Experimental Physics,
Moscow, Russia}
\address[kur]{Kurchatov Institute, Moscow, Russia}
\address[korea]{Korea University, Seoul, South Korea}
\address[ires]{Institut de Recherches Subatomiques, IN2P3-CNRS and
Universit\'e Louis Pasteur, Strasbourg, France}
\address[warsaw]{Institute of Experimental Physics, Warsaw University,
Poland}
\address[zagreb]{Rudjer Boskovic Institute, Zagreb, Croatia}

\begin{abstract}
The $\phi$-meson production cross section is measured for the first
time at a sub-threshold energy of $1.93$ AGeV in
\nuc{58}{Ni}+\nuc{58}{Ni} central collisions.
The $\phi$ data were obtained within the acceptance of the CDC/Barrel
subsystem of FOPI.
For a sample of $4.7\cdot 10^6$ central events, after background
subtraction, $23$ candidates were observed.
Extensive GEANT simulations of the detector performance are shown in a
thorough comparison to the real response, aiming at a good
understanding of the apparatus and at a trustable determination of the
efficiencies, production probability and possible systematic errors.
A filter procedure is elaborated, which is meant to facilitate the
comparison of any theoretical calculation or new data with the current
ones.
How to extrapolate the present value to a $\phi$-meson cross section
in $4\pi$ is also discussed.
This result on pseudo-vector mesons can now be compared to existing
experimental knowledge for the same reaction at the same incident
energy for various outgoing channels, $K^+$ and $K^-$ included.
A significant fraction (at least $20\%$) of the $K^-$-mesons is
originating in the decay of the $\phi$, supporting the statement that
the two channels are strongly correlated.
\end{abstract}

\end{frontmatter}

\section{Introduction}

Relativistic heavy ion collisions offer a unique possibility to probe
experimentally hot and dense nuclear matter: at bombarding energies of
$1-2$ AGeV baryonic densities of $2-3$ times the saturation value
$\rho_0$ at zero temperature are reached, however the condition lasts
merely about $15$ fm/c and the degree of equilibration established in
such a short interval is open to questions~\cite{Rami:00}.
A long standing goal of the study of central reactions in the beam
energy range $1-10$ AGeV is the determination of the compressibility
$\kappa$ of nuclear matter at non-zero temperature and high density.
Unfortunately, almost all the proposed observables sensitive to
$\kappa$ have at some point been discovered unreliable.
Many forms of collective motion (flow~\cite{Reisdorf:97}) were thought
to manifest simply the effect of the compressibility of nuclear
matter, but were later proven to be strongly affected also by the
momentum dependent part of the mean field so that the former one can
not be simply disentangled~\cite{Gale:90,Aichelin:91,Pan:93}
(in ref.~\cite{Pan:93} a reasonable estimate of $165<\kappa<220$ MeV
was derived from the data).
Recently elliptic flow has been proposed as a possible test
quantity~\cite{Danielewicz:98}.
Pion production was
believed~\cite{Stocker:78,Harris:85,Harris:87} to be another good
observable to extract $\kappa$; but the high absorption in nuclear
matter makes it susceptible not only to the compression phase and
maximum density but also to the details of the
expansion~\cite{Bertsch:84,Molitoris:84,Kruse:85}.
Finally, Aichelin and Ko~\cite{Aichelin:85} proposed sub-threshold
$K^+$-production as a promising probe: in fact strangeness is
conserved in strong interactions and $K^+$ absorption is negligible in
nuclear matter (strangeness exchange is present but at equilibrium
should be balanced) and hence the yield is insensitive to the
expansion phase.
Because of the Pauli blocking the direct nucleon-nucleon reaction,
that could happen with the help of the Fermi momentum, is almost
suppressed and at sub-threshold  energies the dominant creation
mechanism is a two step one.
The conclusion has been confirmed experimentally by the KaoS
collaboration measuring the variation in the yield with the number of
participants to the central
collision~\cite{Grosse:93,Miskowiec:94,Senger:99}.
The production probability is then tightly bound to the maximum
density reached in the compression phase.

However such apparently clear situation is obscured by the possibility
of in-medium effects on the $K^+$, $K^-$ and in general meson
properties.
Despite the small alteration of the $K^+$ mass in the nuclear
environment, the experimental determination of the EOS is rendered
difficult~\cite{Hartnack:00}.
Studying the relation between system size and $K^+$-production with
high quality data, in order to distinguish between the in-medium and
fire-ball density (i.e. EOS) contributions, has accumulated an
evidence in favor of a soft equation of state of nuclear matter
($\kappa\approx 200$ MeV)~\cite{Sturm:01}.

Nevertheless the subject of the possible influences of the surrounding
nuclear medium on the meson masses is interesting by its own, with
important implications in
astrophysics~\cite{Brown:88,Brown:94a,Brown:94b,Li:97}.
For example, eventually the $K^-$ condensation in neutron
stars~\cite{Kaplan:86,Nelson:87,Brown:87} could lead to a softening of
the equation of state of nuclear matter above $3$ times the saturation
density $\rho_0$.
The evolution of supernova explosions is than altered causing a core
with $1.5-2$ solar masses to collapse into a black hole rather than to
form a neutron star~\cite{Brown:94a,Brown:94b,Li:97}.

In particular it is in general well accepted that in nuclear matter at
equilibrium increasing temperature and/or density will lower the so
called quark condensate $\langle q\overline{q} \rangle$ thus partially
restoring the chiral symmetry.
However, while the prediction of the consequences on pseudo-scalar
mesons (like the $K^+$ and $K^-$) is robust to different model
frameworks~\cite{Schaffner:97}, the situation appears disputed for
pseudo-vector mesons (like the $\phi$) more exposed to problems in
the theoretical
description~\cite{Brown:91,Hatsuda:92,Kuwabara:95,Klingl:97,Friman:97,Klingl:98}.
Additionally the $\phi$-meson yield is coupled to the sum of the
$K^+$ and $K^-$ mass.
If, as it is currently hypothesized, the reduction of the latter can
be as high as $\approx 100$ MeV for $\rho=3\,\rho_0$, the
corresponding decay width in the nuclear medium is diminished
increasing the absorption probability.
Actually Pisarski and independently Shuryak and
collaborators~\cite{Shuryak:91,Lissauer:91,Shuryak:92} originally
proposed to study the kaon in-medium properties from an enhancement or
suppression of the $\phi$-production probability in heavy ion
collisions, implicitly considering the modification of the $\phi$ mass
itself less important.
Hartnack, Oeschler and Aichelin have suggested that the $K^-$ yield is
almost insensitive to its own in-medium mass, but is instead strongly
linked to the $K^+$ one by the secondary interaction
$K^-B\leftrightarrow\Lambda(\Sigma)\pi$~\cite{Hartnack:02}.
All these arguments indicate the strong interconnection between the
$K^+$, $K^-$ and $\phi$-channels and assess the impossibility of
studying each of them separately.
The present result is the missing piece to complete the already
available knowledge on $K^+$ and
$K^-$-mesons~\cite{Best:97,Menzel:00,Wisniewski:00} for the same
reaction and beam-energy.

Besides, due to $\omega-\phi$ mixing, the $\phi$-meson wave function
is dominated by the $s\overline{s}$ component and so its propagation
in nuclear matter is interesting for complementing the informations
gained from hadrons with open strangeness
($K^+$,$K^-$,$K^{0}_{S}$,$\Lambda$)
~\cite{Ritman:95a,Best:97,Herrmann:99,Wisniewski:00,Finck:01}.

Recently the parameterization of the $NN$ in-medium cross section has
been shown to be the major element of uncertainty for the calculation
of the charged kaons production probability.
As a matter of fact the disagreement between the Giessen
group~\cite{Cassing:99} and Nantes
group~\cite{Aichelin:91,Hartnack:98} transport codes, quite dissimilar
on a physical basis, was found to be marginal, under such respect,
when a common parameterization is supplied~\cite{Hartnack:02}.
It is also comparable to the predicted variation between the
in-medium and no in-medium effect scenarios.

The long awaited measurement of both the elementary processes
$pp\rightarrow K^- X$ \cite{Balewski:96,Balewski:98} and
$pp\rightarrow\phi pp$~\cite{Balestra:99,Balestra:01} near threshold
allows to nail down in a critical region the freedom in the cross
section parameterizations necessary for theoretical models,
consistently reducing ambiguities.

In this work the data from FOPI restricted to the CDC/Barrel subsystem
on $\phi$-meson yield in \nuc{58}{Ni}+\nuc{58}{Ni} at $1.93$ AGeV will
be introduced.
From within the limited experimental acceptance, a $4\pi$
extrapolation can exclusively be done once the shape and
characteristics of the source are assumed.
The channels and thresholds for $\phi$-meson creation are not too
different from that of a $K^+\,K^-$ pair and, as recalled, they are
strongly related: our best estimate of the $\phi/K^-$-ratio for
central collisions will be discussed.
A significant part of the $K^-$, at least $20\%$, seems to arise from
the decay of the $\phi$.
Preliminary hints of sub-threshold $\phi$-meson production have
already been reported using both the CDC/Barrel~\cite{Herrmann:96} and
the Helitron/Forward Wall~\cite{Kotte:00} combinations, here the
emphasis is on the extraction of a quantitative estimate of their
occurrence probability in central collision.

\section{Data analysis}

\subsection{The FOPI detector}

The FOPI detector is a modular system for fixed target experiments at
the SIS accelerator of GSI-Darmstadt.
The four subdetectors (CDC, Barrel, Helitron and
Forward-Wall~\cite{Gobbi:93,Ritman:95b}) are coupled two by two
(CDC/Barrel and Helitron/Forward-Wall) to provide momentum,
energy-loss and time of flight information in the polar angles
$39^\circ < \vartheta_\mathrm{lab}< 133^{\circ}$ and $7^\circ
<\vartheta_\mathrm{lab}< 28^{\circ}$, respectively.
The coverage in the azimuthal angle $\varphi$ is almost complete,
except for the Plastic Barrel.
The reference polar coordinate system of the current work has the beam
direction (i.e. $z$-axis) as opening direction and the target as
origin for the polar angle $\vartheta_\mathrm{lab}$.
The azimuthal angle $\varphi$ lies in the plane orthogonal to the beam
axis containing the target position (i.e. $xy$-plane).

The Barrel and the CDC are operated in a magnetic field of
$0.6$ T produced by a super-conducting coil.
The Helitron is a vector chamber and together with the Forward-Wall is
placed in the fringe field of the solenoid in the forward hemisphere.
In the ongoing analysis the CDC/Barrel subsystem alone is involved and
in what follows we focus on it.

The CDC is a drift chamber of the Jet
type~\cite{Farr:78,Drumm:80,Fischer:86,Heuer:88,Fischer:89,Opal:91}
divided in $16$ sectors with projective geometry for the forward edge
(inner radius $20.7$ cm and length $75$ cm, outer radius $80.1$ cm and
length $192.5$ cm, see table~\ref{tacc}) each equipped with $60$ sense
wires in the median plane.
Such choice has the advantage of allowing a more uniform charge
collection and multiplication field together with a reduction of the
drift length by a factor of $2$, still keeping an equal amount of
electronic channels.
However, the effective readout comes on cost of the left-right
ambiguity for every hit that for tracking purposes needs to
be considered in a pair of spatial positions of which just one
corresponds to the correct localization.
To partly remedy the problem the sectors are tilted by $8^\circ$ with
respect to the ideal radial symmetry around the target so that only
one of the two mirror images points to the latter.
In addition, the sense wires are
staggered~\cite{Farr:78,Drumm:80,Fischer:86,Heuer:88,Fischer:89,Opal:91}
by $\pm$ $200$ $\mu$m about the nominal sense wire plane giving an
independent criterion for the distinction of real and mirror tracks.
The gas mixture chosen is $88\%$ $Ar$, $2\%$ $CH_4$ and $10\%$
iso-butane at atmospheric pressure.

In Jet chambers the transverse position of the hit in the $xy$-plane
is deduced from the cell number and measured drift time, while that
along the $z$-direction by the charge division between the ends of the
wire: the physical basis is quite different for the two coordinates
and also the physical mechanisms imposing the actual limits to the
resolutions are not the same; as a consequence the first is known
much better than the second.
Such feature is taken into account in the processing
algorithm where the hits candidate to form a track are identified
initially in the $xy$-projection of the event~\cite{Drumm:80} requiring
the distance from the nearest neighbor to be less than
$\Delta_{xy\,hit}$ (see table~\ref{tcut}).
In the successive step the position along $z$ of the hits pertaining
to a track is examined and all those further apart from the closest
one of $\Delta_{z\,hit}$ (see table~\ref{tcut}) are rejected as not
belonging to it.
For each particle the transverse momentum component $p_t$ over charge
$Z$ (in units of the elementary charge) can be obtained by the inverse
$k$ of the radius of the arc of circle fitted through the
$xy$-projection of the trajectory, as
\begin{equation}
\frac{p_t}{Z}=\frac{0.3\,B\,\mbox{GeV/c T}^{-1}\mbox{m}^{-1}}{k}=
\frac{0.18\,\mbox{GeV/c m}^{-1}}{k}\;,
\label{pt}
\end{equation}
where in the last expression we substituted $B=0.6$ Tesla for the
magnetic field of FOPI.
From the angle $\vartheta_\mathrm{lab}$, delivered by the straight
line fitted to the $z$-coordinate of the hits, the total momentum $p$
over charge $Z$ and its components are found
\begin{equation}
\frac{p}{Z}=\frac{p_t}{\sin\,\vartheta_\mathrm{lab}}\;.
\label{pperp}
\end{equation}
The error on $p$ is
\begin{eqnarray}
\sigma_p^2/p^2 & = &\sigma^2_{p_t}/{p_t}^2+\cot^2(\vartheta_{lab})\,
\sigma^2_{\vartheta_{lab}}\nonumber\\
& = & \sigma^2_k/k^2+\cot^2(\vartheta_{lab})\,
\sigma^2_{\vartheta_{lab}}
\end{eqnarray}
where the two autonomous contributions are apparent: $\sigma_k$,
coming from the determination of the radius of curvature of the track
in the transverse plane and,
$\sigma_{\vartheta_\mathrm{lab}}$, from the reconstruction of the
emission angle $\vartheta_\mathrm{lab}$.
In the CDC the physical origins of $\sigma_k$ and
$\sigma_{\vartheta_\mathrm{lab}}$ are quite dissimilar as we have
already discussed.
The condition $2/k>R_\mathrm{min}$ for the trajectory to enter the CDC
can be reexpressed with eq.~(\ref{pt}) as $p_t > 0.019$ GeV/c.

The particle identification is based on a double technique allowing a
redundancy very powerful to reject the background.
Once the momentum is known, the charge over mass ratio can be
determined (and hence the species assigned) if the velocity is
measured.

\begin{figure}[H]
\centering\mbox{\epsfig{file=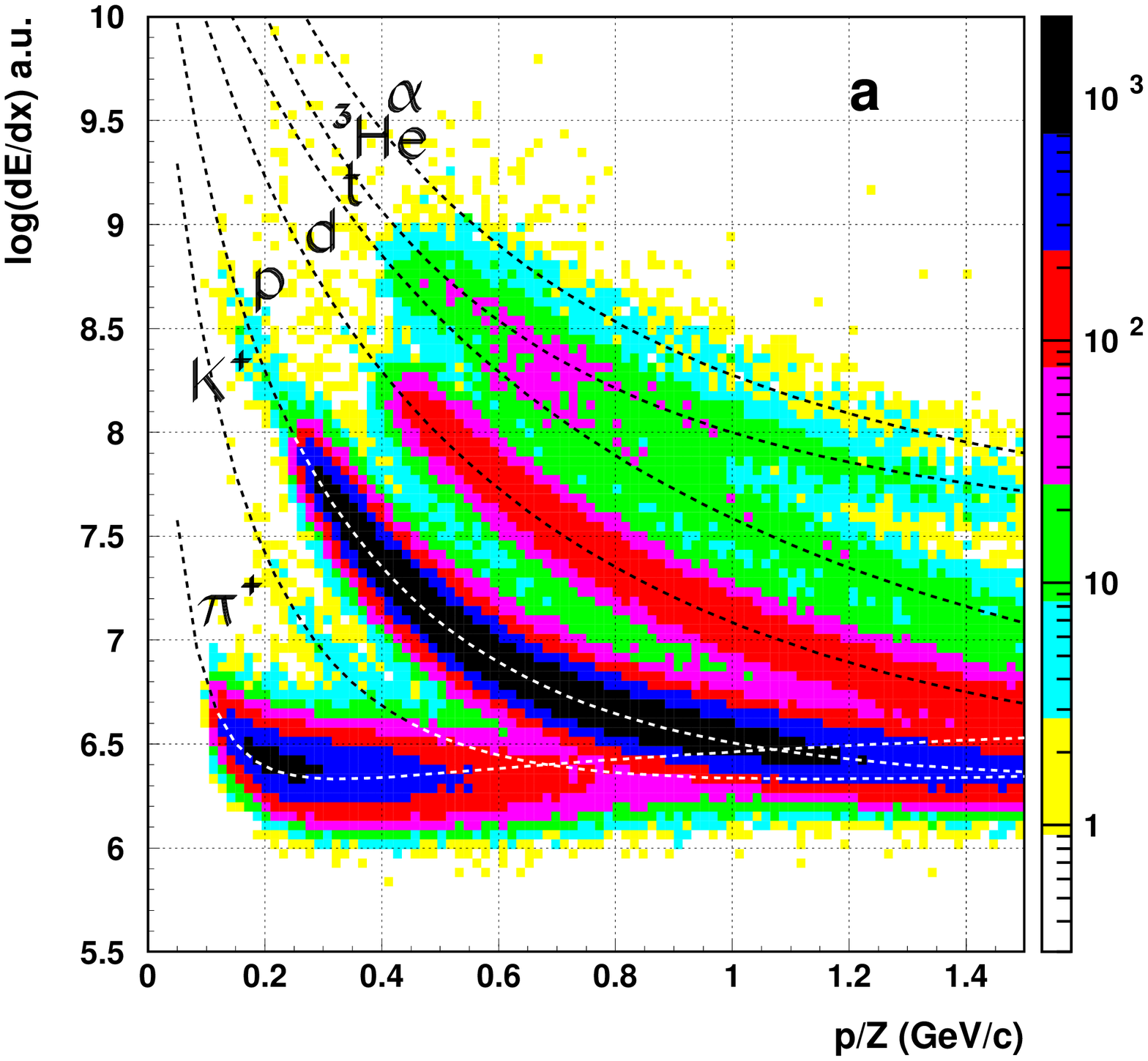,width=.5\textwidth}\epsfig{file=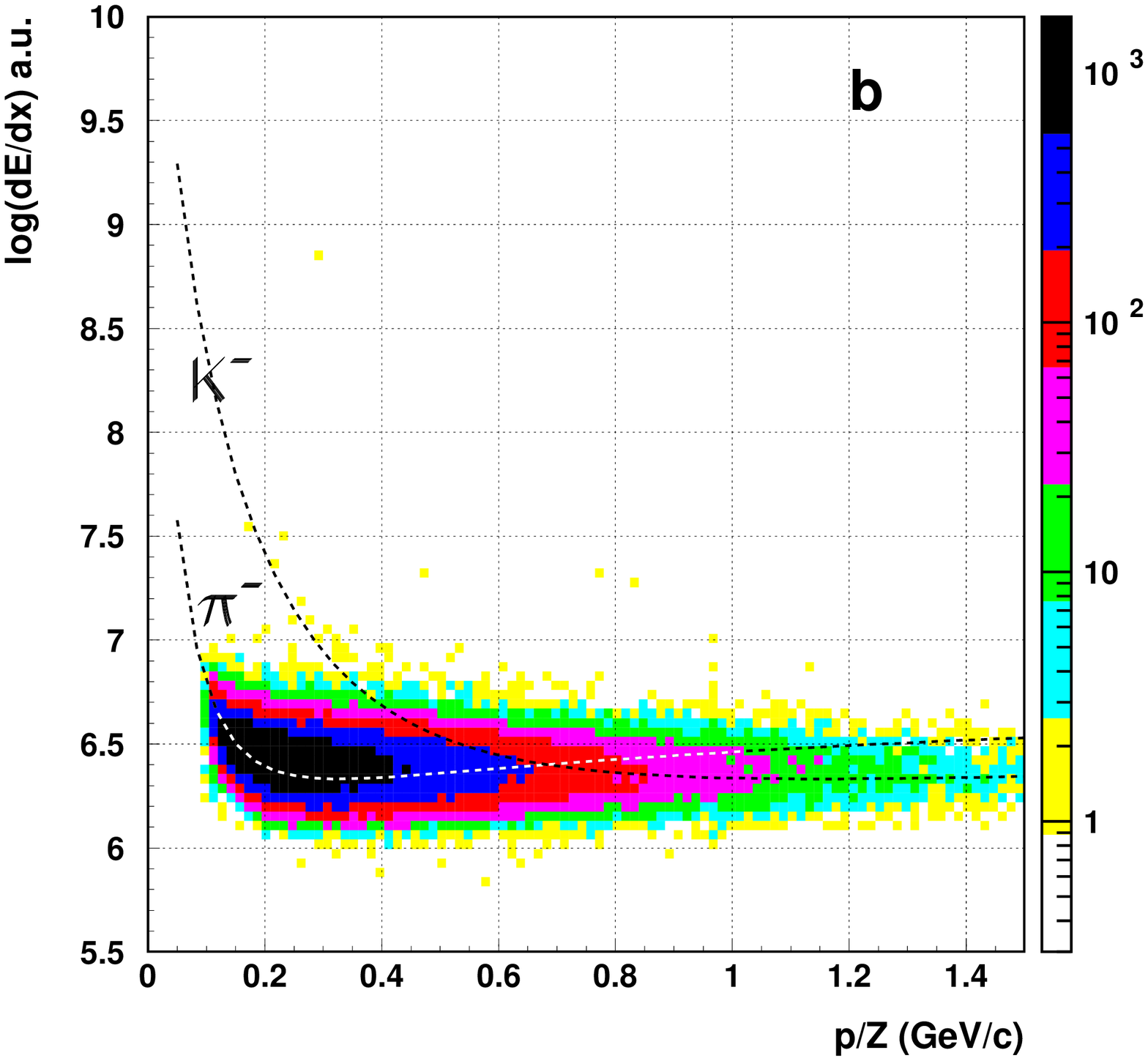,width=.5\textwidth}}
\centering\mbox{\epsfig{file=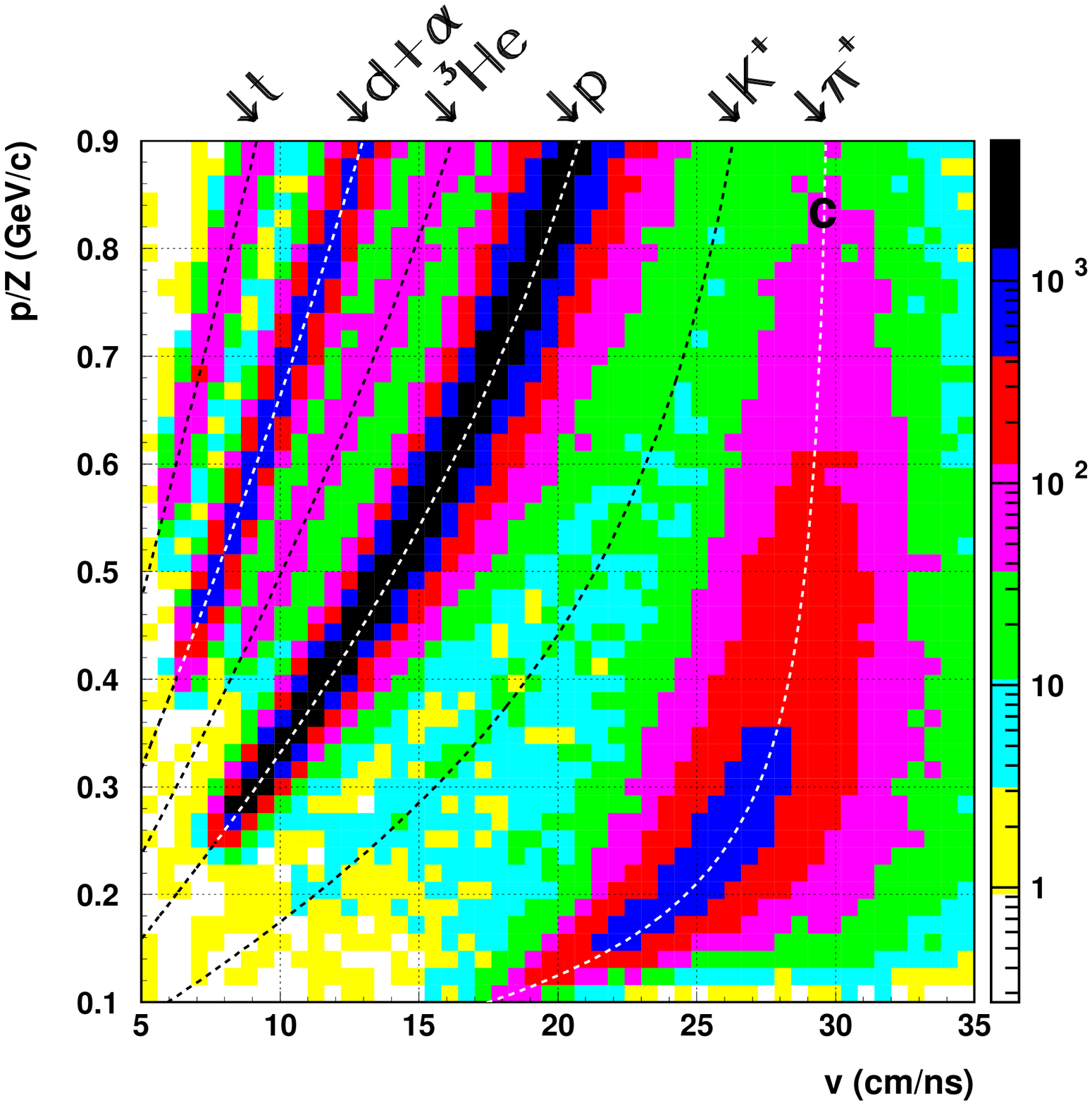,width=.5\textwidth}\epsfig{file=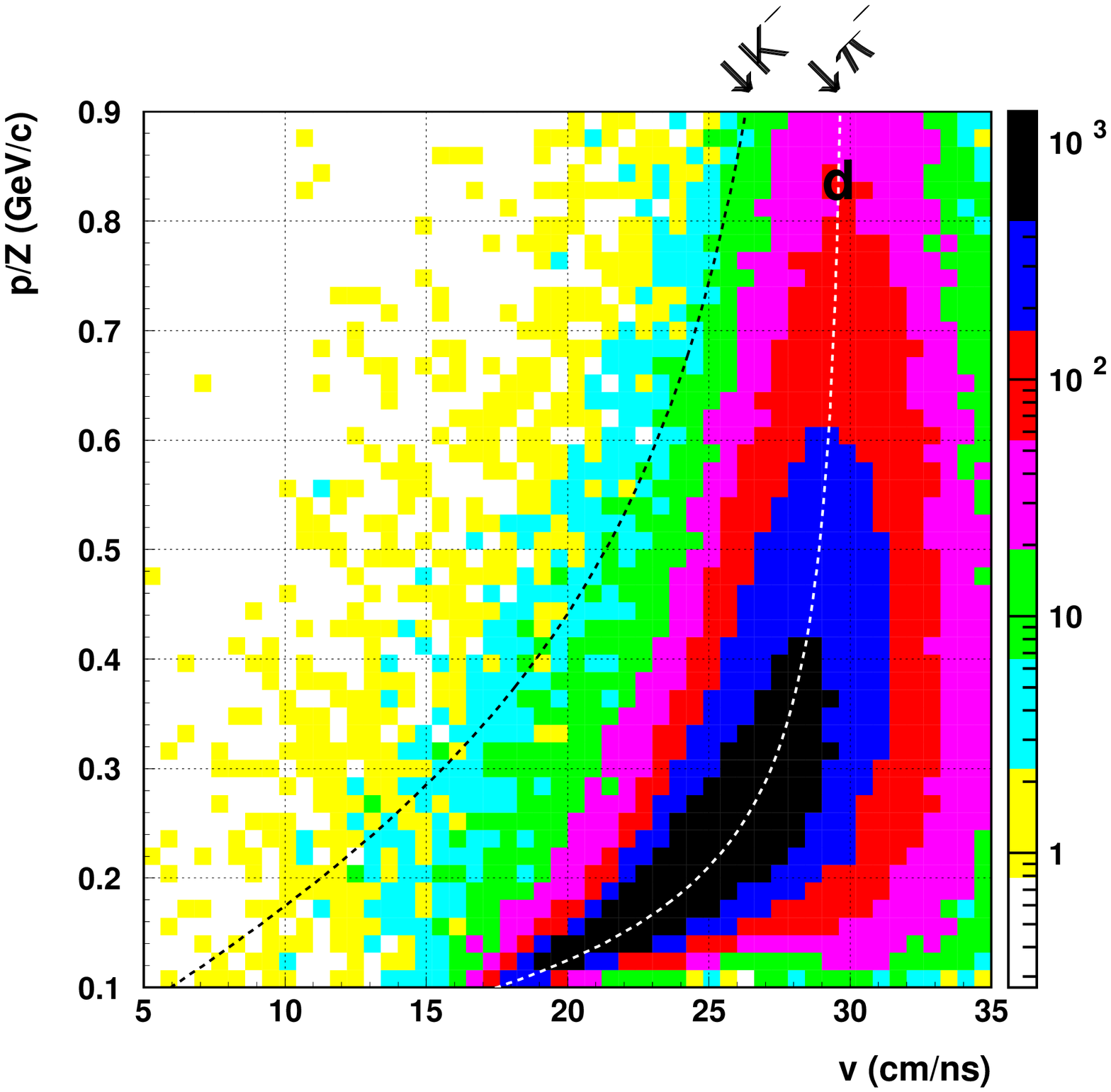,width=.5\textwidth}}
\caption{Upper panels: logarithm of the truncated mean
energy-loss $dE/dx$ versus momentum $p/Z$ for positively (panel~a) and
for negatively (panel~b) charged particles reaching the Barrel.
The dashed lines represent the empirical parameterizations used
for the energy-loss versus momentum in the $Cmass$ reconstruction
algorithm; the constants chosen are the same as those of the data
analysis for the extraction of the $\phi$-candidates.
Lower panels: momentum $p/Z$ versus velocity $v$ for positively
(panel~c) and for negatively (panel~d) charged particles.
The dashed lines represent the relativistic momentum versus velocity
relation with the nominal masses.}
\label{pid}
\end{figure}

Exploiting the dependence of the mean energy-loss on the velocity, and
not on the momentum, of the particle sweeping through the detection
medium (i.e. the CDC gas), using an empirical parameterization, a mass
named $Cmass$ can be extracted.
In fig.~\ref{pid} (upper two panels)  the truncated mean
energy-loss~\cite{Drumm:80,Heuer:88} (after correction for the
$\vartheta$ angle of incidence in each cell evaluated from the track
fit) is shown versus momentum: the branches corresponding to
$\pi$, $K$ and $p$ are discernible and separable.
Since when the Barrel is reached the track length inside the CDC and
hence the number of sampling points for the energy-loss distribution
are constrained, only such cases are considered in fig.~\ref{pid},
besides it will be the final requirement of the present analysis.
The continuous curves, also drawn, refer to the empirical
parameterization defining the regions where the various species can
be identified.

An uncorrelated information on the mass is derived for the ones that
survive up to the Barrel by the time of flight and the length of the
extrapolated trajectory fitted to the CDC hits (a condition is imposed
on the difference $\Delta_\varphi$ in $\varphi$-angle and $\Delta_z$
in $z$-coordinate between the measured and expected impact position in
the Barrel, see table~\ref{tcut}).
In this case the velocity is determined directly.
The Barrel consists of $180$ bars of scintillator of width $3.25$ cm,
depth $4.05$ cm and length $240$ cm arranged in $30$ modules (the
space for $2$ further ones is occupied by the CDC support structure)
disposed in a cylindrical fashion at a radius of
$R_\mathrm{Bar}=111.5$ cm coaxially with the $z$-axis and covering the
$\vartheta_\mathrm{lab}$ angles reported in table~\ref{tacc}.
Due to the missing modules and to the inter-module and inter-bar
spacings also the azimuthal coverage is incomplete.
The scintillators are readout with photo-multipliers located outside
the FOPI magnet through long light guides.
The condition $2/k>R_\mathrm{Bar}$ for the Barrel to be reached can be
reexpressed with eq.~(\ref{pt}) as $p_t> 0.10$ GeV/c.
In table~\ref{tacc} we have summarized the acceptances of the CDC and
Barrel subdetectors of FOPI.
In fig.~\ref{pid} (lower two panels) the velocity is displayed versus
momentum and again the branches corresponding to $\pi$, $K$ and $p$
are distinguishable.
The continuous curves are the well known relativistic expression in
the nominal cases. 
A mass named $Bmass$ is calculated as
\begin{equation}
Bmass = \frac{p}{c\,\beta}\,\sqrt{1-\beta^{2}}\;.
\end{equation}

The $Cmass$ is limited at high momentum from the relativistic rise of
the energy-loss as a function of $p$: above $\approx 0.6$ GeV/c the
pion and kaon region merge and contamination is unavoidable
(see fig.~\ref{pid}~a and~b).

Also $Bmass$ is limited, for kaons, below $\approx 0.6$ GeV/c by the
finite time resolution of the Barrel ($\approx 400$ ps) joined to the
divergence in the relativistic momentum-velocity relation (see
fig.~\ref{pid}~c).

The advantage of the redundant simultaneous measurement of both
$Cmass$ and $Bmass$, is illustrated in fig.~\ref{bmcm} representing
one mass versus the other: a structure corresponding to the $K^+$ is
clearly visible, the $K^-$ are less well separated owing to their
small abundance. 
To perceive an idea of the difficulty of such task, let us remind that
in the acceptance for $1$ proton there are roughly $\approx 10^{-1}$
pions, $\approx 10^{-3}$ $K^+$ and $\approx 10^{-5}$ $K^-$.
The superior quality of $Bmass$ over $Cmass$ can be deduced from the
deformation of the distribution along the latter.
For this reason they will not be treated alike in the final
identification cuts.

\subsection{Event selection}
\label{evtse}

\begin{figure}[H]
\centering\mbox{\epsfig{file=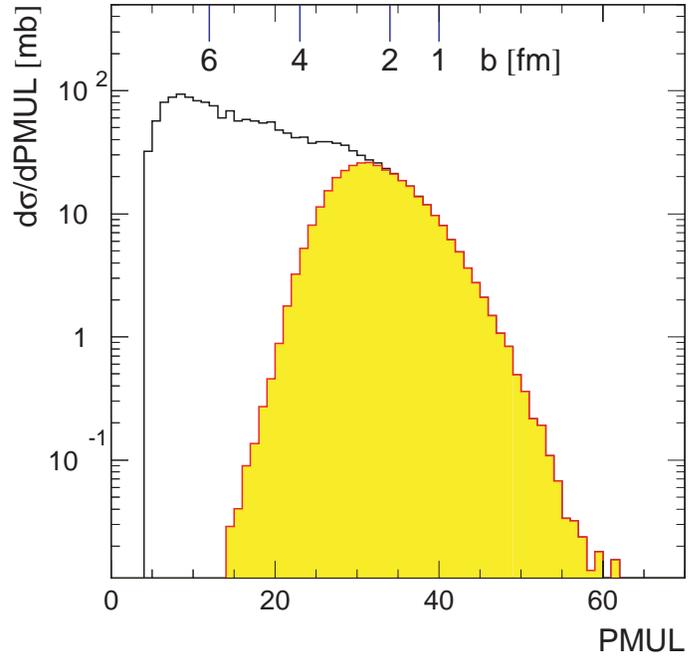,width=.7\textwidth}}
\caption{Charged particle multiplicity (PMUL) distributions
corresponding to the Minimum Bias trigger (empty) and the Central
trigger (shaded).
The upper scale in impact parameter $b$ is derived with the assumption
of a sharp cut-off geometrical model.}
\label{trig}
\end{figure}

The online event centrality selection is based on the detected
charged particle multiplicity in the Forward-Wall, $M$, obtained with
fast analog electronics~\cite{Gobbi:93}.
A minimum bias trigger is defined when more than $4$ are found, and a
central one when $M>M_c$.
During event acquisition, the quantity $M_c$ is set so that the ratio
between the minimum bias and central triggers corresponds to about
$10\%$~\cite{Gobbi:93}.
In particular we focus on the $7.7\cdot 10^6$ central collisions
collected with the FOPI detector in $1995$ for the
\nuc{58}{Ni}+\nuc{58}{Ni} reaction at $1.93$ AGeV beam kinetic
energy.
After imposing constraints on the energy deposited in the start
counter, for the purpose of optimizing the time resolution, and on the
primary interaction vertex position in both the transverse $xy$-plane
and $z$-direction, for the purpose of reducing background from
interactions not produced in the target, a sample of $4.7\cdot 10^6$
events is left.
In the more accurate offline reprocessing a certain jitter is observed
in the multiplicity threshold for the central trigger caused by the
non-ideal behavior of the analog electronics with which it is
implemented (see fig.~\ref{trig}).
The similarity of a selection on minimum multiplicity $M$ to an upper
limit on impact parameter $b(M)$ is straightforward in the sharp
cut-off hypothesis assuming a monotonic relation between $M$ and $b$.
Adopting a simple geometrical model the central collision condition
cited through this work can be seen to be equivalent to
$0<b<\approx 3.3$ fm or to the most central $\sigma_c\approx 350$ mb
(i.e. $12\%$) of the geometrical reaction cross section
$\sigma_\mathrm{geo}\approx 2.9$ b~\cite{Menzel:00}.
A rather simple comparison is possible for the yield per central event
between our experimental finding and transport codes once the
the prediction is studied as a function of centrality.
For the theoretical result $P_\mathrm{th}(b)$ an appropriate mean
$\langle P_\mathrm{th}\rangle$ over $b$ in the range $0<b<3.3$ fm is
to be computed taking as a weight the ratio of the differential
reaction cross section $d\sigma_R/d\,b$ to the integrated one
$\sigma_c$
\begin{equation}
\langle P_\mathrm{th} \rangle = \frac{2}{b_\mathrm{max}^2}
\int_{0}^{b_\mathrm{max}} P_\mathrm{th}(b)\,b\,db\;,
\label{mpb}
\end{equation}
where $b_\mathrm{max}$ is $3.3$ fm in the case of FOPI.

\subsection{$\Phi$-meson identification}
\label{phiid}

The $\phi$-meson is identified through its decay channel in a $K^+$
and a $K^-$ ($BR=49.1\%$).
The charged kaons are pre-selected mainly with a window in both
$Cmass$ and $Bmass$ and an upper momentum limit.
A data sub-sample is generated with kinematic and other quantities.
For calculating the invariant mass of the couple, the common nominal
value of the two species is used.
The combinatorial background is reconstructed via the event-mixing
technique~\cite{Berger:77,Jancso:77}: at the same time an independent
data sub-sample is generated regrouping in turn each kaon from the
current event together with the opposite charged partner from a stack
of depth $20$ of previous ones.
The latter are sorted with respect to the amount of particles detected
in the Forward-Wall (i.e. impact parameter, see sec.~\ref{evtse}).
The stack can be filled with central collisions where a $K^+$ or a
$K^-$ is pre-identified or where both a $K^+$ and a $K^-$ are
pre-identified.
The momentum vectors from the stack are rotated around the beam axis
so that their reaction plane has the right current orientation.
The population size in the event-mixing is easily made much larger
($2$ orders of magnitude) than in the data, in such fashion the
statistical error contribution to the outcome for the
$\phi$-candidates is negligible.

To reject mirror or erroneous tracks, three general cuts to ensure
good quality are imposed: on $d_{0}$ (the minimum distance in the
transverse $xy$-plane of the trajectory from the reconstructed event
vertex position), on $z_{0}$ (the displacement from the target of the
point where the helix crosses the $z$-axis) and on $n_{hit}$ (number
of hits): $|d_{0}|<0.5$ cm, $|z_{0}|<10$ cm and $n_{hits}>30$ (see
table~\ref{tcut}).
A difference between $d_0$ and $z_0$ should be noted: while the target
thickness is small and the initial collision $z$-coordinate is well
defined, the corresponding statement is not true for the beam spot
transverse dimension and $d_0$ refers to the $xy$-position extracted
from the data.
By the way, the acceptance windows on $d_{0}$ and $z_{0}$ also help to
reduce the background of pions coming from secondary vertices.
All tracks fulfilling the $d_{0}$ and $z_{0}$ selections are then
refitted forcing their origin in the event estimated first interaction
point.
This new helix is then the base for calculating the momentum and for
performing the extrapolation to the Barrel necessary to evaluate the
matching condition and the length of flight.
In the present situation the procedure can not rise any problem since
all the $\phi$-mesons are decaying almost immediately outside the
fireball (with $c\tau = 44.5$ fm and $p=1$ GeV/c the mean length
traveled is $43.7$ fm) and hence they are all well inside the target.
It is of course not accomplished in those studies, like on
$\Lambda$-baryon production, in which detection of a vertex apart from
the primary reaction one has to be
exploited~\cite{Ritman:95a,Finck:01}.

\begin{figure}[H]
\centering\mbox{\epsfig{file=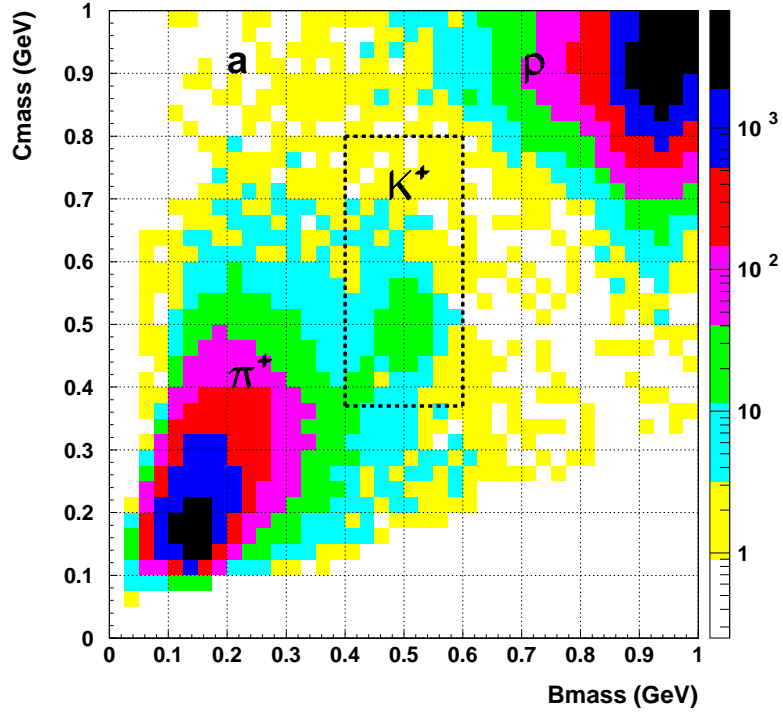,width=.75\textwidth}}
\centering\mbox{\epsfig{file=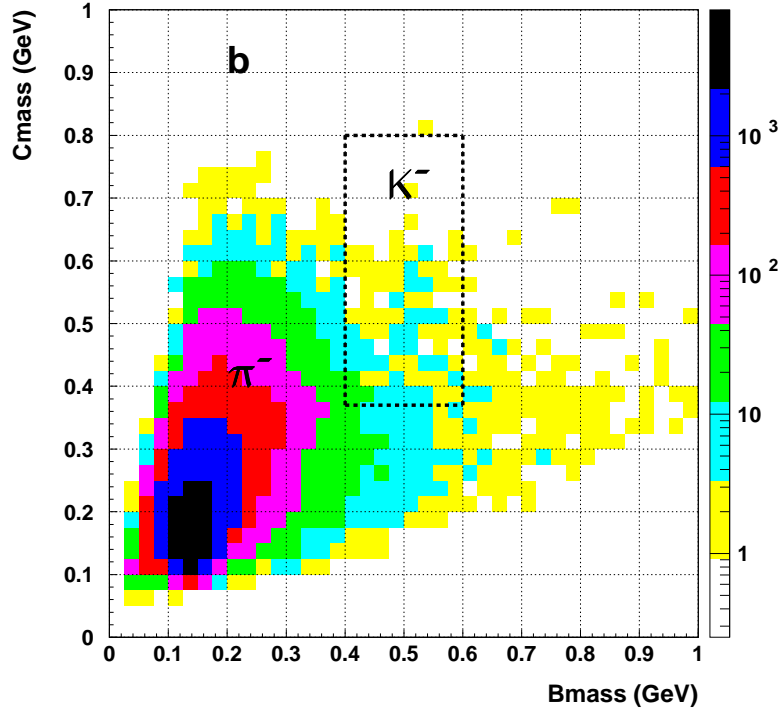,width=.75\textwidth}}
\caption{$Cmass$ versus $Bmass$ for positively (panel~a) and
negatively (panel~b) charged particles with momentum less then
$0.6$ GeV/c.
The area enclosed by the dashed line corresponds to the kaon
identification cuts used in the $\phi$-meson analysis (see
table~\ref{tcut}).}
\label{bmcm}
\end{figure}

\begin{figure}[H]
\centering\mbox{\epsfig{file=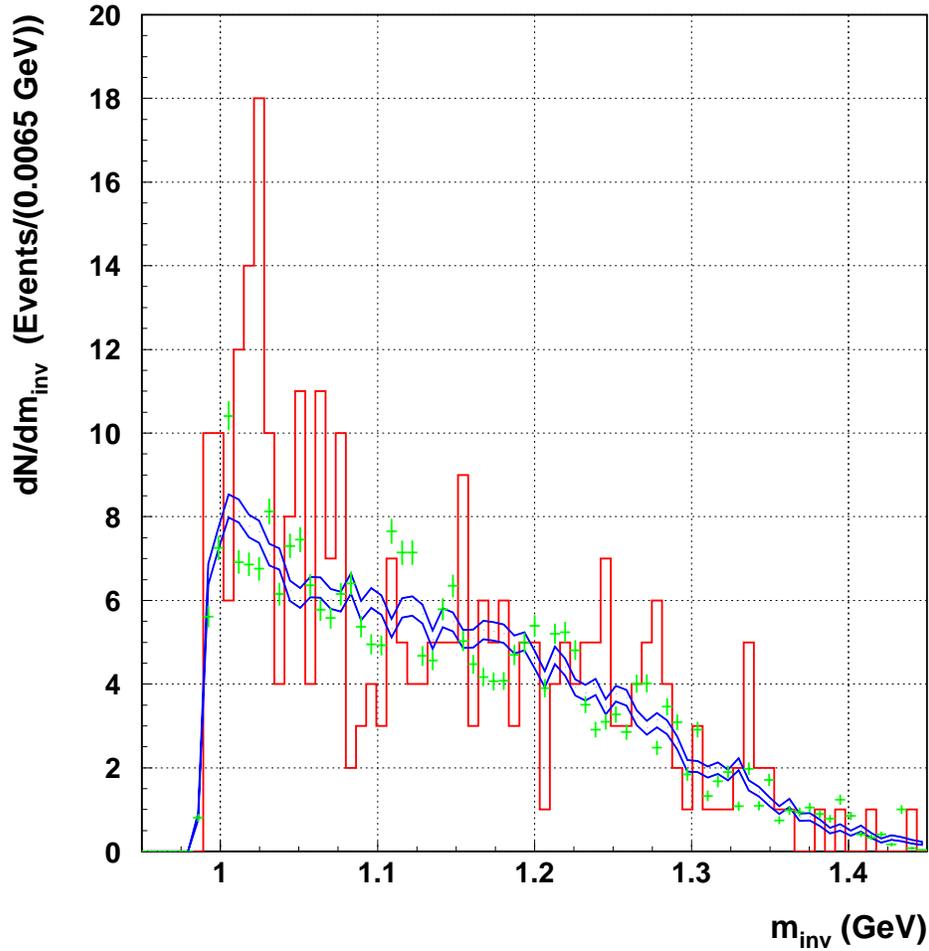,width=1.\textwidth}}
\caption{Invariant mass spectra for identified $K^+\,K^-$ pairs with
the final cuts (see table~\ref{tcut}).
The histogram is the signal; the two lines enclose one standard
deviation from the background reconstructed with the event-mixing
technique when the mixing stack is filled with \nuc{}{Ni}+\nuc{}{Ni}
collisions containing either a $K^+$ or a $K^-$-meson; the error bars
refer to the background reconstructed when the mixing stack is filled
with reactions containing both a $K^+$ and a $K^-$.}
\label{minv}
\end{figure}

The final cuts (see table~\ref{tcut}) $p_\mathrm{lab}<0.6$ GeV/c,
$Cmass>0.37$ GeV (for pion background suppression, see
fig.~\ref{bmcm}~a and~b), $Cmass<0.8$ GeV (for proton background
suppression, see fig.~\ref{bmcm}~a)  and $|Bmass-0.494|<0.1$ GeV are
then applied for particles of both charges to the two sub-samples
(data and event-mixing) and the $K^+\,K^-$ invariant mass
distributions are constructed (as shown in fig.~\ref{minv}).
The normalization is provided integrating the spectra aside from the
$\phi$-peak ($m_\mathrm{inv}>1.06$ GeV/c).
The combinatorial background drawn with the lines enclosing one
standard deviation in fig.~\ref{minv} is obtained filling the mixing
stack with events containing either a $K^+$ or a $K^-$, the other,
represented with the standard error bars, when the mixing stack is
filled from events containing a $K^+$ and a $K^-$.
The first is much more regular and will be employed in the subsequent
analysis.
The second has still instabilities of non statistical origin
indicating that, although the whole sample is just $30\%$ smaller, the
poorer class from which it is drawn (i.e. the same element of the
stack is reused more often) makes the destruction of the correlations
harder.
It can be considered solely for estimating a kind of maximum
systematic error due to self correlation induced by the limited
population adopted for mixing (such source would not be important in
an analysis with much more candidates).

After subtraction, the number of events in the $\phi$-peak (defined
from the end-points at $1.01$ and $1.03$ GeV where the remainder of
the signal and the background becomes negative) with its statistical
error is of $23\pm7$.
In a previous analysis~\cite{Mangiarotti:00}, done with older
calibrations, a result of $22$ was found.
From this difference and from the difference obtained subtracting
both backgrounds a systematic error of $\pm2$ or $9\%$ is estimated.

\section{Simulation}

The $(23\pm7\pm2)$ $\phi$-mesons given in sec.~\ref{phiid}, are of
little utility for a comparison with future experiments or with
theoretical calculations.
An accurate determination of the detection efficiency by simulation is
needed to deduce a cross section within the phase space coverage (or
in $4\pi$).

In the present section we will always assume that all $\phi$-mesons
are decaying through the $K^+\,K^-$-channel.
The appropriate branching ratio will be included in sec.~\ref{pmpy}
for deducing the production probability.

The overall efficiency $\epsilon_\mathrm{tot}$ to measure in our
device such particle out of $4\pi$ can be splitted into two terms
\begin{equation}
\epsilon_\mathrm{tot}=\epsilon_\mathrm{max}\,\epsilon_\mathrm{det}\;,
\label{eff}
\end{equation}
where $\epsilon_\mathrm{max}$ is defined as the efficiency of
measuring a $\phi$ within the geometrical and momentum acceptance of a
"perfect" apparatus plus the effect of kaon decay in flight during
their path to the Barrel and $\epsilon_\mathrm{det}$ contains the
further reduction introduced by detector limitations and by the
constraints necessary for a clean identification (vertex, matching and
mass cuts).

Information on the $\phi$ distribution in $4\pi$ can not be obtained
in the current effort because the phase space region covered by the
CDC/Barrel subsystem around mid-rapidity is small.
For an extrapolation into the full solid angle of the cross section
extracted in the acceptance, one has to rely upon a $4\pi$ shape.

The advantage of the factorization~(\ref{eff}) is that while
$\epsilon_\mathrm{max}$ depends essentially on the initial phase space
distribution in $4\pi$, $\epsilon_\mathrm{det}$ does not, within a
good approximation.
Such case will be demonstrated to hold in a representative example in
sec.~\ref{tdeefg} and~\ref{feff} where the "effective" temperature of
a one source thermal model is varied.
Thus the comparison between our experiment and new ones or different
calculations is greatly facilitated.
For any elaborated study delivering the entire $\phi$ population, it
is sufficient to apply a simple "filter" procedure (see
appendix~\ref{filter}).
This takes into account the detector geometry and momentum cut as
well as the kaon decay, neglecting first all complicated details of
the apparatus performance, which can be considered separately through
the constant coefficient $\epsilon_\mathrm{det}$, determined once for
all with the necessary knowledge of FOPI in sec.~\ref{feff}.

\subsection{The CDC/Barrel maximum acceptance}
\label{maxac}

In the study developed here the simple hypothesis of a thermal source
controlled by only one parameter (i.e. the temperature $T$) is
adopted.
A more refined model would be the Siemens and
Rasmussen~\cite{Siemens:79} formula, containing a temperature
$T_\mathrm{SR} = 92$ MeV and a collective expansion with velocity
$\beta_\mathrm{flow} = 0.3$.
The whole idea behind this description of a central collision is that
$T_\mathrm{SR}$ and $\beta_\mathrm{flow}$ are common for all hadrons
and so their values can be supplied by a previous study done with the
FOPI detector about $p$, $d$ and $\pi^-$ from the same
\nuc{58}{Ni}+\nuc{58}{Ni} reaction at the same beam-energy of $1.93$
AGeV~\cite{Hong:98}.
However, as long as just the $\phi$-meson is concerned and within the
necessary accuracies, the source can be approximated with a pure
thermal one with $T = 127$ MeV.
The influence of the dynamics can then be reabsorbed  in the
temperature $T$, that becomes distinct from the real one $T_{SR}$
(and species associated).
Transport model calculations generally follow a similar trend: the
shape of the phase space distributions can be considered thermal
within the present statistical and systematic errors, but the
effective temperatures are lower, typically $70-90$ MeV.
We can than in rough approximation reduce our uncertainty on the
source to a thermal source where the temperature is unknown by a wide
margin.

To investigate the change in CDC and Barrel acceptances with
temperature $T$ in the interval $130-70$ MeV, four GEANT simulations
(see table~\ref{teff}) were undertaken.
However the detector response description and event reconstruction
stages are not involved in the extraction of such estimates and solely
the FOPI geometry implementation and particle transport part of the
code were exploited.
The results are summarized in table~\ref{teff} with and without kaon
decay enabled.
The column corresponding to $\epsilon_\mathrm{max}$ is the third from
the end: a total excursion with temperature by a factor of $\approx 4$
is found.
This clearly illustrates the impossibility of giving one cumulative
number and the necessity of the decomposition~(\ref{eff}) in order to
allow a comparison with the predictions.

\begin{figure}[H]
\centering\mbox{\epsfig{file=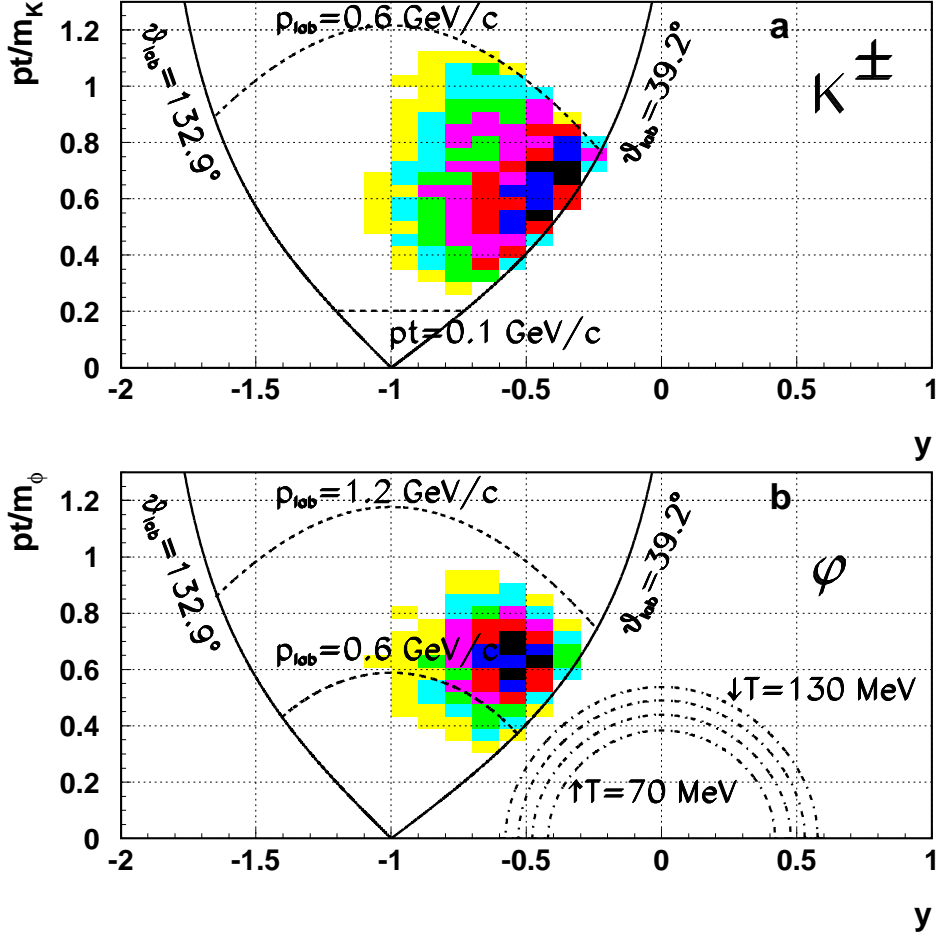,width=1.\textwidth}}
\caption{Transverse momentum over mass $p_t/m$ versus rapidity $y$
plots for the original $\phi$-meson (panel~b) and for the $K^+$ and
$K^-$-mesons produced in its decay (panel~a).
The GEANT simulation with a source temperature of $130$ MeV and
including the decay in flight of the kaons is used.
The rapidity is renormalized such that $-1$ is the target, $+1$ is the
projectile rapidity and $0$ is mid-rapidity, where the $\phi$ source is
located.
The continuous and dashed lines in panel~a and~b illustrate the
acceptance limits of the CDC/Barrel combination (compare with
table~\ref{tacc}).
The dot-dashed curves in panel~b represent the most probable momenta
for a relativistic Maxwell-Boltzmann distribution of emitted $\phi$
centered at $y=0$ and with a temperature of $130$, $110$, $90$ and $70$
MeV, respectively.}
\label{ypt}
\end{figure}

It is important to realize that the smallness of
$\epsilon_\mathrm{max}$ derives from the localization of the phase
space covered by the CDC/Barrel around the target rapidities.
In fig.~\ref{ypt}~a the $p_t/m$ versus normalized rapidity $y$ plot is
shown for the $K^+$ and $K^-$-mesons reaching the Barrel and
originating in the decay of the $\phi$ emitted by a thermal source
with $130$ MeV temperature.
As can be seen, the relevant cuts are the Barrel forward
$\vartheta_\mathrm{lab}$ angle limit of $39.2^\circ$ and the maximum
momentum allowance of $0.6$ GeV/c (see table~\ref{tacc}).
In fig.~\ref{ypt}~b the $p_t/m\div y$ representation is used for the
primary quantities of the $\phi$-mesons whose $K^+$ and $K^-$ decayed
into the range of acceptance.
Its maximum is still well away from the mid-rapidity region.
The dot-dashed curves, also in panel~b, correspond to the most
probable momentum for a relativistic Maxwell-Boltzmann $\phi$ emission
law centered at $y=0$.
The portion of the diagram accessible with the CDC/Barrel subdetector
combination is outside this line indicating that exclusively the tail
of the distribution is probed.
This becomes even more pronounced if the source temperature is
lowered, indicating why $\epsilon_\mathrm{max}$ is reduced under such
conditions.

\subsection{The detection efficiency estimation from GEANT}
\label{tdeefg}

As already mentioned $\epsilon_\mathrm{det}$ contains the adjoined
contribution to the efficiency, in the phase space acceptance and
after kaon decay, of the detector behavior.
For its determination a precise reproduction of the apparatus
response is necessary.
The current section is devoted to the discussion of the quality of the
environment implemented in the simulation.
This is important to judge upon the reliability of the conclusions
derived from the latter.

For our study we started from the well proven software package GEANT
(version~$3.21$ distributed with CERNlib version $99$) written at
CERN~\cite{Geant:321}.
In particular this frame comprises very powerful routines for the
description of the detector geometry (only the CDC, Barrel and the
magnet, return iron yoke, CDC/Barrel support structures and
target passive volumes have been implemented for the ongoing attempt).
Once the material of each volume and its characteristics have been
specified the standard package routines are capable of calculating the
particle transport in the magnetic field throughout the detector
including the decay with the appropriate lifetimes and branching
ratios and the interactions with the penetrated media.
Each decay product can then in turn be transported and let
decay until they are all outside the interesting regions.
If a volume is specified as sensitive (e.g. the CDC gas or the Barrel
scintillator) hits are generated with the corresponding positions and
energy-loss values.

The remaining parts under the user responsibility (except the
geometrical and material descriptions) are the event generator and the
hit processing routine necessary to build the final output.

With the thermal source a class of simulations, indicated in the
following as ``single-$\phi$'', has been run.
However the realistic track multiplicity of the full nuclear
\nuc{}{Ni}+\nuc{}{Ni} event, originating the $\phi$-meson, will
certainly reduce the efficiency.
In fact, due to the low granularity of the Barrel, one or more of the
$180$ scintillator bars can be crossed by multiple particles in the
same event.
In such case a sole improper position is recovered and both hits are
discarded in the CDC/Barrel matching phase.
If one of them corresponds to a kaon from the $\phi$ decay, the $\phi$
itself is lost.
The CDC behavior too is not expected to be sampled correctly under
the simplified situation with just two kaons not involving noise hits
together with intersecting tracks and their mirror images.
Each of the preceding reasons of inefficiency is affected by the
distribution of the emitted $\phi$-meson azimuthal angle with respect
to the reaction plane where trajectories tend to concentrate.
To take into account as much as possible the effect of the environment
surrounding the paths of the $K^+$ and $K^-$-mesons created by the
$\phi$ decay, an other class of simulations, named in the following
as ``embedded-$\phi$'', was accomplished.
In it, to the thermally generated $\phi$ a whole IQMD
\nuc{}{Ni}+\nuc{}{Ni} central collision ($b=0$) was superimposed.
The orientation of the reaction plane, with respect to the $\phi$
azimuthal emission angle, was changed at random.
We remind that the version of IQMD~\cite{Hartnack:98} employed
contains no kaons, but only pions, nucleons and heavier fragments.
In the following we will concentrate on single-$\phi$ Monte-Carlo
where the simpler condition allows a detailed analysis of the
stability of the code adopted, a clean comparison with the data and
a reasonable estimate of the systematic errors.
The embedded-$\phi$ simulation is cited to quantitatively evaluate the
additional inefficiency coming from the nuclear environment
surrounding the $\phi$.

In the hit processing routines a description of the front-end
electronics has been embodied with the associated noise.
For the CDC to derive from the information provided by GEANT (position
and energy-loss) the contents of the $100$ Mhz Flash ADC memories,
four steps are necessary.
First, for all tracks in a sector the charge corresponding to the
given energy-loss is distributed with fluctuations along the path
within each individual drift cell between electron-ion clusters.
The latter are afterwards propagated under the position dependent
influence of the electric and magnetic fields~\cite{Drumm:80} up to
the sense wire.
For every cluster a generic pulse shape at each of the ends is also
delivered.
Second, all the pulses arriving at one specific readout channel are
added together time bin wise without regarding the particle that
produced them (in such way a cross-talk between different tracks in a
sector is introduced).
Third, the noise is added in an uncorrelated manner, separately for
each wire, and in a correlated one, injecting charge simultaneously
into all wires of one sector (as it was found to occur in the
experiment).
In both cases a Gaussian distribution is assumed.
Finally, the sum is digitized with a 8 bit nonlinear (10 bits
effective) ADC accuracy.
For the Barrel the time and energy digitized information is derived on
the basis of a constant propagation velocity and attenuation length in
the scintillator bars and light guides.
A Gaussian smearing is added both in time and pulse height to include
experimental resolutions.
Double hits in the same scintillator bar are treated in a fashion that
the fastest signal reaching the photomultiplier defines the time tag
(independently for both ends) and the pulse heights are summed.
Now the events are in a form compatible to be fed as input to the
tracking and reconstruction code with which data are elaborated.

\subsubsection{The basic input resolutions as compared to the
experiment}
\label{tbiractte}

\begin{figure}[H]
\centering\mbox{\epsfig{file=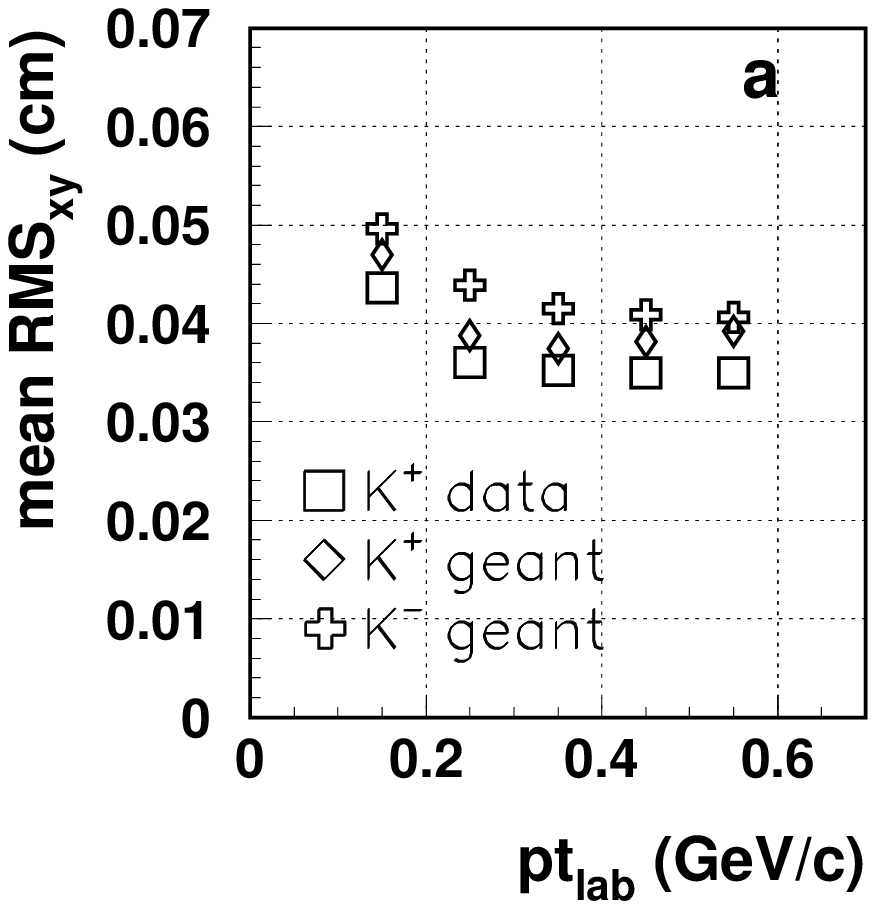,width=.5\textwidth}}
\centering\mbox{\epsfig{file=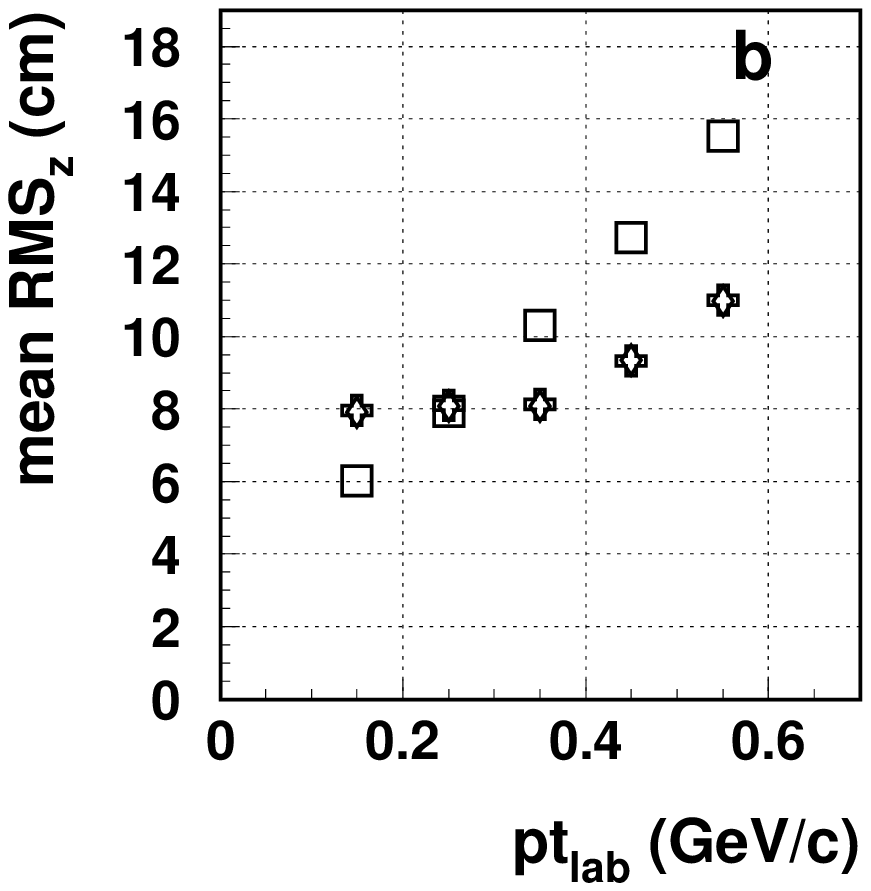,width=.5\textwidth}}
\centering\mbox{\epsfig{file=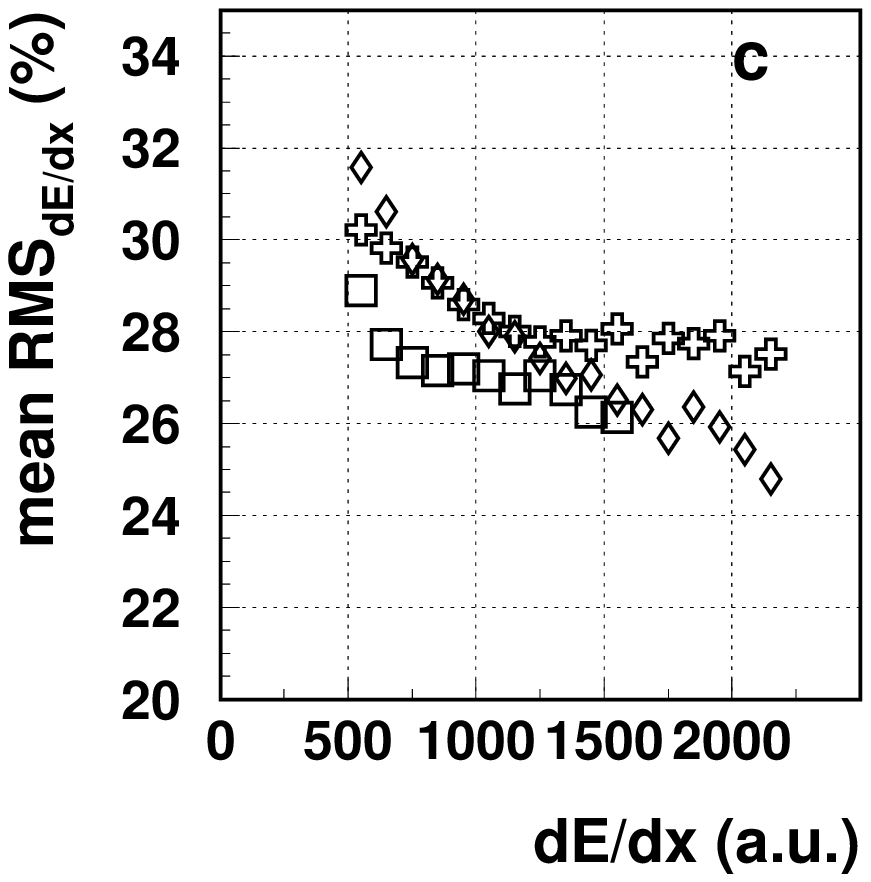,width=.5\textwidth}}
\caption{Results obtained from the CDC digitization routine for the
mean rms $\mathrm{RMS}_{xy}$ of the track fit residuals in the
transverse plane (panel~a), the mean rms $\mathrm{RMS}_z$ of the track
fit residuals in the $z$-coordinate (panel~b) and the mean rms
$\mathrm{RMS}_{dE/dx}$ of the truncated energy-loss distribution
(panel~c).
Note that in panel~b $K^+$ and $K^-$-mesons from the simulation lie on
top of each other.
The range in energy-loss of the data ($500-1600$ a.u.) corresponds to
kaons with momenta $\approx 0.1<p<0.6$ GeV/c.}
\label{gcdc}
\end{figure}

Since the characteristics of the detection system can not be probed by
a well defined beam hitting into the acceptance of the CDC, the
performance of the Monte-Carlo digitization routines has to be
monitored and tuned from the response to usual tracks.
For those, of course, the real transverse momentum $p_t$ or
$\vartheta_\mathrm{lab}$ angle are not known.
To remain as close as possible to the basic physical resolutions
connected to how the chamber operates we have decided to compare the
mean rms of the track fit residuals in the transverse plane
$\mathrm{RMS}_{xy}$, the mean rms of the track fit residuals in the
$z$-coordinate $\mathrm{RMS}_{z}$ and the mean rms of the truncated
energy-loss distribution $\mathrm{RMS}_{dE/dx}$.
The results are displayed in fig.~\ref{gcdc} for $K^+$ and
$K^-$-mesons from the single-$\phi$ simulation and $K^+$ alone from
the \nuc{}{Ni}+\nuc{}{Ni} collisions due to limitations in $K^-$
statistics.

For the data a set of identified kaons, defined as in the $\phi$-meson
analysis, was selected (see table~\ref{tcut}).
It has to be stressed that at a fixed $p$ the energy-loss for unlike
particles is not the same and the CDC behavior is correspondingly
changed; so not all of them together, but exclusively $K^+$-mesons,
should be compared with the calculation.

In the case of the Monte-Carlo only kaons known to have reached the
Barrel sensitive volumes during the GEANT transport phase and with
generated momentum less than $0.6$ GeV/c were considered.
Because at most two tracks per event are present, the position of
the primary collision vertex in the transverse $xy$-plane can not be
pinpointed as in the data; rather the refit is always forced to the
origin where the initial position of all the reaction products is
located.
As default $d_0$ and $z_0$ are taken from the free fit.
Such procedure, adopted here and in what follows, is expected to
reproduce best the experimental situation where the number of CDC
tracks is far greater ($\approx 30$) and hence the determination of
the event vertex $xy$-coordinates is accurate.
Note that the true conditions correspond more to the embedded-$\phi$
scenario.

To gain statistics a $130$ MeV thermal source was used and the kaon
decay inhibited.
A comparison with a temperature of $70$ MeV or with the kaon decay
enabled has exhibited negligible differences below the systematic ones
between data and simulation (this is the case also for the observables
studied in the next section).

The agreement is quite satisfactory for $\mathrm{RMS}_{xy}$ and
$\mathrm{RMS}_{dE/dx}$ (see fig.~\ref{gcdc}~a and c).
There are indications that the increase of $\mathrm{RMS}_z$ with $p_t$
is too weak in the simulation (see fig.~\ref{gcdc}~b).
The stronger rise in the real detector is probably to be ascribed
to a form of common high frequency noise induced in all wires, known
to be relevant in this type of chambers~\cite{Heuer:88}, but not
handled properly in the CDC digitization routine developed for GEANT.
In fact while its repercussions are relatively negligible for
$\mathrm{RMS}_{xy}$, being the transverse position deduced from the
drift time, they are rather important for $\mathrm{RMS}_z$ where the
coordinate is determined from the charge division.

The discrepancies between the values of $\mathrm{RMS}_{xy}$ and
$\mathrm{RMS}_z$ in the Monte-Carlo and the data are small with
respect to the much larger cuts $\Delta_{xy\,hit}$ and
$\Delta_{z\,hit}$ controlling the tracker behavior and so no
readjustment in particular of the latter was done for the analysis of
the simulation (see table~\ref{tcut}).

Once $\mathrm{RMS}_{xy}$ and $\mathrm{RMS}_z$ have been fixed, the CDC
accuracy in transverse momentum $\mathrm{RMS}_{p_t}$ and in
$\vartheta_\mathrm{lab}$ angle $\mathrm{RMS}_\vartheta$ can be
investigated with the Monte-Carlo incorporating the description of the
detector response.
Energy-loss, energy-loss straggling and multiple scattering in the
passive and active volumes of the CDC/Barrel combination were
accounted profiting of the description of the physical processes
embodied in the GEANT software package~\cite{Geant:321}.
The results are summarized in fig.~\ref{gptthe} both for the mean
systematic shifts and the resolution of the reconstructed quantities.

For $p_t$ a maximum systematic variation of $\pm 1\%$ is seen.
The low $p_t$ distortion in fig.~\ref{gptthe}~a is to be attributed to
the energy-loss in the target and chamber entrance materials, as is
proven by the simulation where these passive volumes were removed.
The dissimilar behavior between positive and negative particles in the
high $p_t$ region (see fig.~\ref{gptthe}~a) is due to the sectors
tilting that introduces a systematic asymmetry.
In fig.~\ref{gptthe}~b the $p_t$ resolution is also compared with what
is expected from the analytical formulas of
Gluckstern~\cite{Gluckstern:63,Blum:93} considering the multiple
scattering in the CDC gas (dashed line) and the transverse hit
position measurement error $\mathrm{RMS}_{xy}$ of $400$ $\mu$m, as
deduced from fig.~\ref{gcdc}~a, and under the vertex constraint
(dot-dashed line).
The quadratic addition of the two contributions (continuous line)
is quite near to what is obtained from the Monte-Carlo environment
without target and chamber entrance materials, after tracking.
It can be concluded that the multiple scattering in the chamber gas as
well as in the passive volumes is the dominant limitation in the
extraction of transverse momentum through the entire $p_t$ range of
interest.
In the real experiment, if trajectories need to be merged between
adjacent sectors, the situation is expected to be worsened by other
problems, not included in the digitization routines.

For the $\vartheta_\mathrm{lab}$ angle, the systematic effects can be
neglected as compared to the corresponding accuracy.
The worsening of $\mathrm{RMS}_\vartheta$ toward $90^\circ$ can be
understood as a $z$-coordinate constant resolution (not function of
$z$ itself) which translates into an uncertainty in
$\vartheta_\mathrm{lab}$ angle increasing toward
$\vartheta_\mathrm{lab}=90^\circ$.

The performance of the Barrel is less critical to model as it contains
basically nothing else but the time resolution (being the position
determined by the time difference) and the outcome will be supplied in
the next section in terms of the final quantities.

\subsubsection{The shapes of the distributions associated to the
selection cuts} 
\label{tsotdattsc}

The comparison and tuning of the basic resolutions $\mathrm{RMS}_{xy}$,
$\mathrm{RMS}_z$ and $\mathrm{RMS}_{dE/dx}$ is a prerequisite, but for
the estimation of the detection efficiency the more complex derived
observables, necessary for kaon selection in the $\phi$-meson
production evaluation, have to be studied: $d_0$, $z_0$, CDC/Barrel
matching in $z$-coordinate $\Delta_z$ and in $\varphi$-angle
$\Delta_\varphi$, $Cmass$ and $Bmass$.
For each of these, attention should be devoted to both the mean and
the width.
In all cases the $K^+$ and $K^-$ from the single-$\phi$ and the $K^+$
from the \nuc{}{Ni}+\nuc{}{Ni} FOPI data have been chosen.
In the simulation, again the sample of particles known to have reached
the Barrel sensitive volumes during the GEANT transport phase was
always taken (even for variables like $d_0$) and the initial momentum
was limited to $0.6$ GeV/c.
The values of the data refer to kaons identified with the cuts of the
$\phi$ analysis (see table~\ref{tcut}).
Only the constraint relevant for the quantity under consideration was
removed to see the shape of the distribution.
A Gaussian fit was found satisfactory in all cases except the
CDC-Barrel matching in $\varphi$-angle, where the influence of the
scintillator bar finite width is present.
Let us discuss all of them in turn.

\begin{figure}[H]
\centering\mbox{\epsfig{file=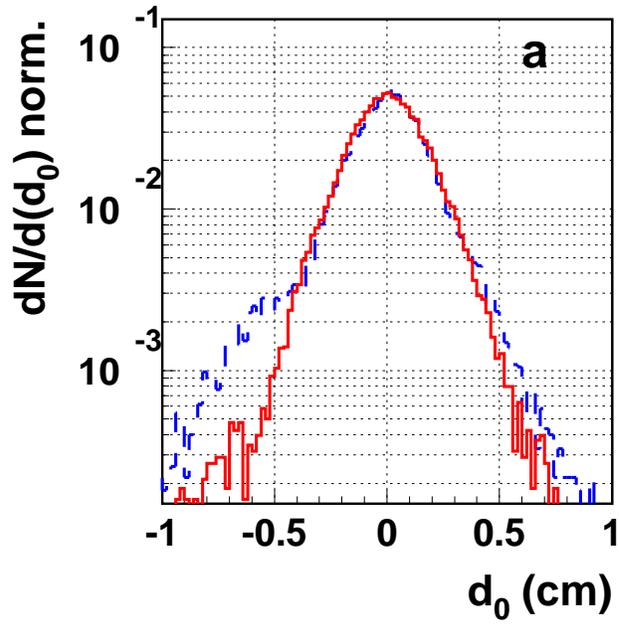,width=.7\textwidth}}
\centering\mbox{\epsfig{file=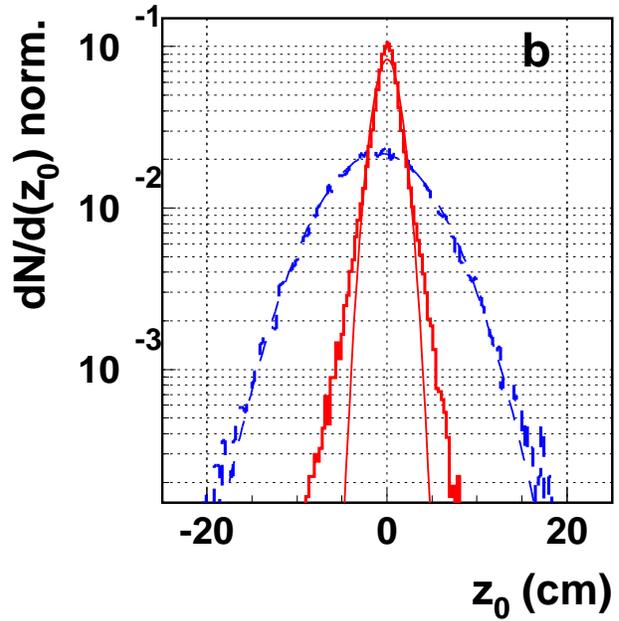,width=.7\textwidth}}
\caption{Comparison of distributions in $d_0$ (panel~a) and $z_0$
(panel~b) of $K^+$-meson tracks.
The solid histograms are from the data and the dashed ones from the
simulation.
In panel~b a Gaussian fit is also shown.}
\label{gver}
\end{figure}

For $d_0$ and $z_0$ the results of the comparison are given in
fig.~\ref{gver}~a and b, respectively.
The mean and width of $d_0$ are in good agreement, the sole
discrepancy is in the bump starting at $-0.5$ cm in the simulation.
The feature is connected to the problem that parts of the mirror
tracks are sometimes also misinterpreted as real new ones.
In the Monte-Carlo it happens in $0.07\%$ and $3\%$ of the cases for
the images of $K^+$ and $K^-$-mesons, respectively (with a statistics
of 75000 tracks).
The already mentioned tilting of the sectors renders the situation
asymmetric and the missidentification is not equal for positive and
negative particles.
The bump is not visible in the data for a simple reason of statistics:
the numbers of $K^+$ and $K^-$ in the acceptance are the same when
exclusively the $\phi$-meson decay is assumed as a source, however in
the experiment the first is greater than the second by $2$ orders of
magnitude~\cite{Wisniewski:00} and so the incomplete rejection of the
image of negative kaon trajectories at the level of a few percent
contributes a $10^{-4}$ background for positive ones.

The width of the calculated $z_0$ distribution is about $4$ times as
wide as measured.
The reason is an underestimation of the correlated electronic noise
pick-up that adds pulse height coherently into all preamplifier
channels on one side of the CDC.
Due to the projective geometry of the chamber this causes a strong
distortion of the angle while the intercept with the beam axis is
relatively stable.
Since we operate with very wide windows on $z_0$ no attempt was made
to fine tune its behavior.
The cut at $\pm10$ cm in the true conditions does not discard any good
track but reduces the background from fake ones and from trajectories
originated by secondary vertices, correspondingly the permitted range
in the processing of GEANT output was opened to $\pm20$ cm.

\begin{figure}[H]
\centering\mbox{\epsfig{file=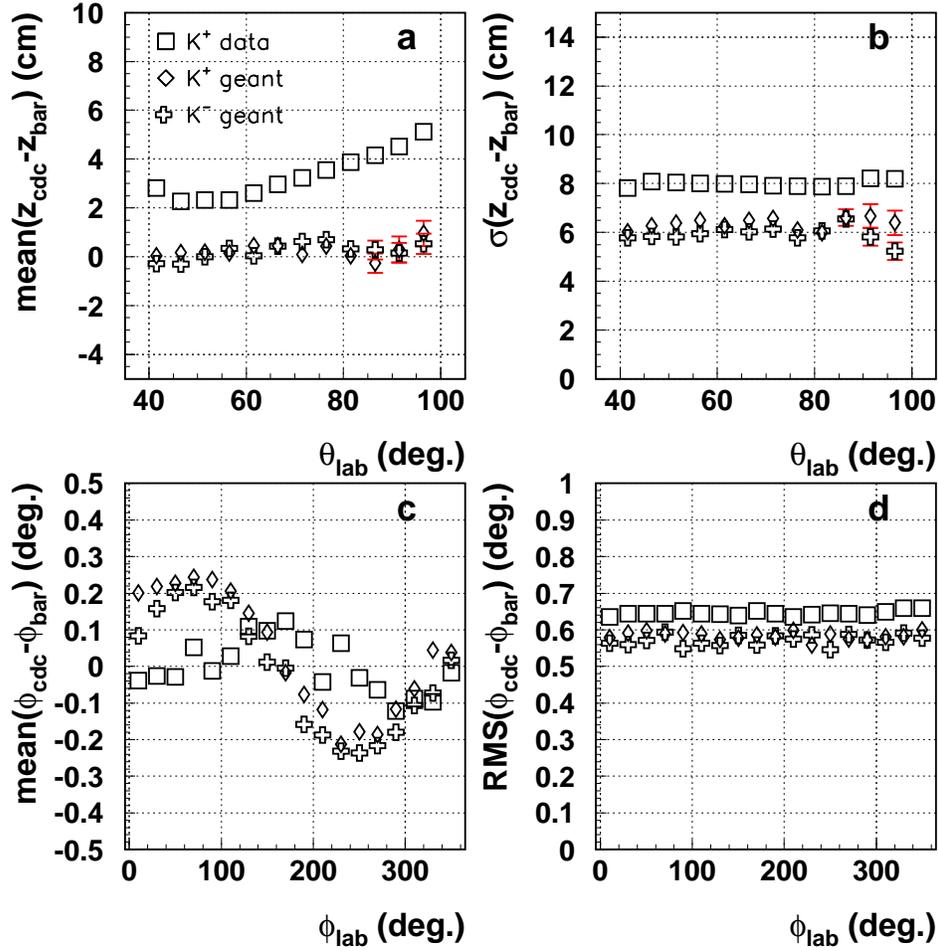,width=1.\textwidth}}
\caption{Comparison of the means and widths for CDC/Barrel matching in
$z$-coordinate and $\varphi$-angle ($\Delta_z$ and $\Delta_\varphi$,
respectively) between $K^+$ and $K^-$-mesons in the simulation and
$K^+$ in the data.}
\label{gmatch}
\end{figure}

The expected shift of the mean as well as the precision for the
CDC-Barrel difference in both $z$-coordinate and $\varphi$-angle
($\Delta_\varphi$ and $\Delta_z$, respectively) are gathered together
with those of the data in fig.~\ref{gmatch}.
For $\Delta_z$, the systematic displacement from $0$ of the latter
(see fig.~\ref{gmatch}~a) is motivated by the difficulty of
calibrating the longitudinal coordinate in the real experiment,
however it is less important than the corresponding standard deviation
by at least a factor of $\approx 2$.
The accuracy in $\Delta_z$ is somewhat better in the Monte-Carlo (see
fig.~\ref{gmatch}~b), so the limits of the allowed region in its
analysis were rescaled in the ratio $6/8$ from $\pm25$ cm to $\pm19$
cm (see table~\ref{tcut}).
In the simulation, the systematic oscillating shift in $\varphi$-angle
matching is imputable to a small inconsistency in the description of
the CDC geometry, it is anyhow $\approx 3$ times less than the
resolution.
The rms of $\Delta_\varphi$ is dominated by the shadow of the
scintillator bar physical size and is hence easily reproduced.

\begin{figure}[H]
\centering\mbox{\epsfig{file=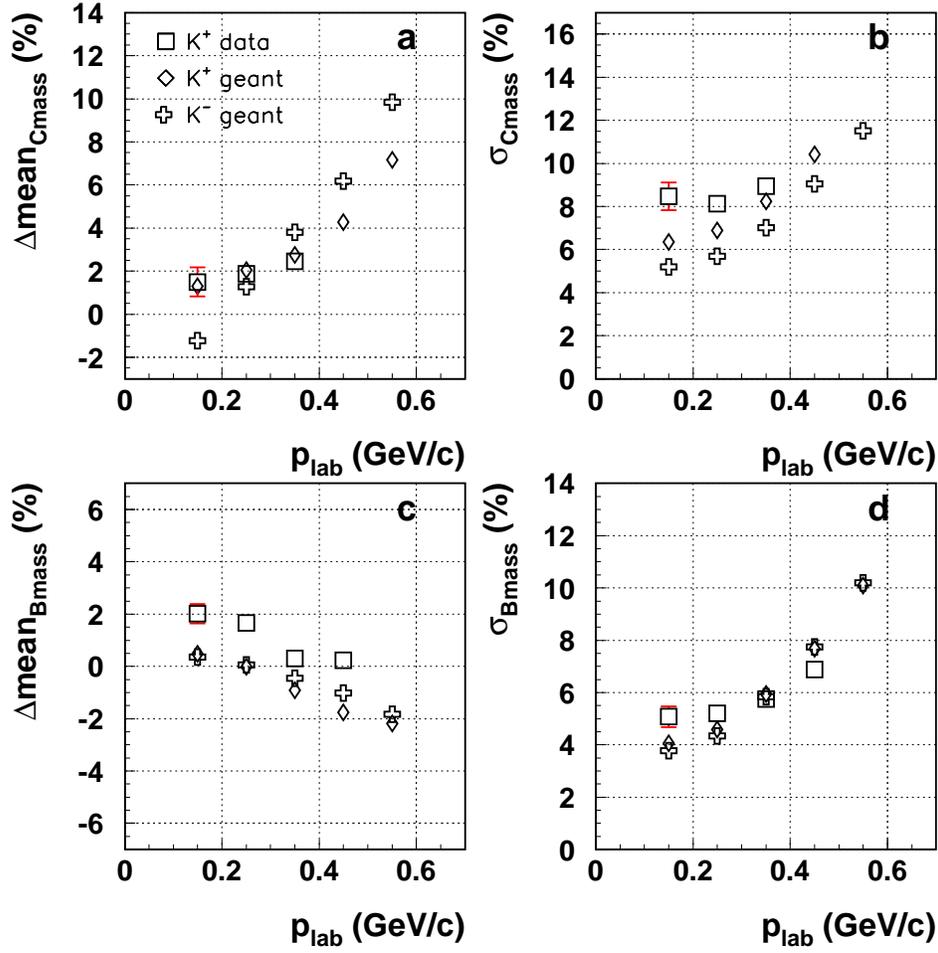,width=1.\textwidth}}
\caption{Comparison of means and widths of the non dimensional
quantities $Cmass/m_K-1$ and $Bmass/m_K-1$ between $K^+$ and
$K^-$-mesons in the simulation and $K^+$ in the data.}
\label{gmass}
\end{figure}

The comparisons of mean position and width of the  $Cmass$ and $Bmass$
distributions are shown in fig.~\ref{gmass} in momentum bins of 0.1
GeV/c up to 0.6 GeV/c.
As dimensionless fractions $Cmass/m_K-1$ and $Bmass/m_K-1$, where
$m_K$ is the nominal charged kaon mass, were preferred to the bare
$Cmass$ and $Bmass$, respectively.
With this choice both the mean and the standard deviation $\sigma$ can
be expressed in percentage.
All trends are similar.
For $Cmass$ the systematic shift (see fig.~\ref{gmass}~a) reflects how
well the employed empirical parameterization is adapted to the
energy-loss versus momentum dependence.
It is a function of the calibration constants and was tailored to
follow as closely as possible the experimental one (obtained in a far
more complex environment with many particle species and hence subject
to more restrictions than in the Monte-Carlo).
A residual uncertainty at the highest momenta is possible between the
$10\%$ of the simulation and the $5\%$ that could be extrapolated from
the data with a straight line.
The width in $Cmass$ (see fig.~\ref{gmass}~b) is dominated by the
$\mathrm{RMS}_{dE/dx}$, found to be in good accordance (see
fig.~\ref{gcdc}~c), it is then not surprising that the measurements
are appropriately modeled.
However substantial extrapolation of the experimental points is
necessary to compare the resolution to the simulation at the
highest allowed momenta and as a consequence a margin of variation
between the $8\%$ constant value of the first and the $13\%$ of the
second is possible.
For $Bmass$ less freedom is available since no empirical
parameterization is needed and the discrepancy in mean position (see
fig.~\ref{gmass}~c) could not be reconciliated, it is anyhow at
worst of $2\%$, much less than the associated precision.
Its accuracy is dominated by the scintillator Barrel time of flight
resolution (found to be $400$ ps), again good agreement exists with
the data (see fig.~\ref{gmass}~d), but substantial extension is needed
at the highest momentum and a discordance remains from the $10\%$
predicted at $0.6$ GeV/c and the $8\%$ that could be extrapolated with
a straight line from the latter.

The estimated possible maximum discrepancies in mean or width (for
all relevant cases the distributions are Gaussian both in the data and
the simulation so that the standard deviation $\sigma$ can be adopted)
discussed in detail in this paragraph are summarized in
table~\ref{tsys}.
They will be the base to estimate the associated systematic
uncertainty in the extraction of $\epsilon_\mathrm{det}$ (see
appendix~\ref{syserr}).

The accepted ranges in the analysis of the GEANT and real results are
collected in table~\ref{tcut}.

In the Monte-Carlo the contribution of each cut can be sorted out
within the statistical accuracy of $\pm1\%$.
The exclusion of hit multiplicities lower then $30$ causes no decrease
and a reduction of $97\%$ on the negative and positive track numbers,
respectively, owing to the more frequent rejection of $K^-$ mirror
images, that has already been discussed in connection with the $d_0$
distribution shape.
The efficiency of the $d_0$ selection itself, for the same problem,
is also charge asymmetric being $95\%$ for positive and $98\%$ for
negative tracks.
The $z_0$ and $Cmass$ allowed windows do not leave out anything.
The requirements on $\Delta_\varphi$ and $\Delta_z$ are strongly
correlated, so that once the $\varphi$-angles selection is imposed,
not much more is eliminated by the one on the $z$-coordinate
difference.
They will be lumped together as a single condition in the following.
The match, $Bmass$ and momentum cut all have an efficiency near
$97-98\%$.
The previous three figures have then to be squared to take into
account the positive and negative kaon identification in every event
to know the impact on the pair (i.e. $\phi$) population.
The $3\%$ missing after matching can not be ascribed to the
$\Delta_\varphi$ and $\Delta_z$ cuts, that correspond to more than $3$
$\sigma$, but rather to the scintillator efficiency.
Although for a path impinging along the normal to the bar entrance
surface, the depth of $4.05$ cm is more than enough to generate a
light signal above threshold in $100\%$ of the cases, it is
nevertheless not true if the trajectories are curved by the magnetic
field, and edge effects become important.
It should be noticed that the width is $3.25$ cm, to be compared with
the already mentioned depth.

A contamination at the level of $(1.3\pm 0.2)\%$ of the
$K^+\,K^-$-couples with both momenta below $0.6$ GeV/c from the part
of the kaon spectrum above $0.6$ GeV/c is found with the present
resolution of the CDC in the simulation.

\subsection{Final efficiency}
\label{feff}

The position and width of the $K^+\,K^-$ invariant mass peak are
determined only by the $p_t$ and $\vartheta_\mathrm{lab}$ angle
systematic shifts and resolutions because, as already mentioned, once
a particle is selected as a kaon candidate the nominal mass is
assumed.
If the peaks are fitted with Gaussian in the region of interest
($1.01-1.03$ GeV, as utilized for extracting the number of
$\phi$-mesons in sec.~\ref{phiid}) the mean is found to be
$1.022\pm0.003$ GeV (data) and $1.020\pm0.003$ GeV (simulation) and
the standard deviation $\sigma$ to be $5.8\pm1.4$ MeV (data) and
$4.7\pm0.3$ (simulation).
However Lorentzian tails outside the fit region are distinguishable
solely in the Monte-Carlo; the experimental spectrum is too much
altered by the lack of statistics and the presence of the background.
To grasp an idea of how looks the comparison to the $\phi$-meson
intrinsic $\Gamma$ of $4.43$ MeV, the FWHM of the Gaussians can be
used: they are a factor of $\approx 3$ larger.
The agreement between the real and expected invariant mass resolution
confirms that the broadening should be attributed to the CDC momentum
uncertainty.
Since the first and second moments are compatible in the measurements
and the simulation, for calculating the fraction of identified $\phi$
an identical binning is performed and the counts in the same $4$
consecutive channels are added together.
From the Monte-Carlo it can be estimated that $(10\pm 3)\%$ of the
events are lost in such a way.

Adopting the previous procedure the last two columns of
table~\ref{teff} have been obtained.
The detection efficiency of the CDC/Barrel subsystem of FOPI for the
$\phi$, independently of the temperature in the considered range
within one standard deviation, amounts to
\begin{equation}
\epsilon_\mathrm{det} = 70\%\pm 1\%\;.
\label{edet}
\end{equation}
The total efficiency with a thermal source of a temperature of $130$
MeV is expected to be
\begin{equation}
\epsilon_\mathrm{tot}= (0.95\pm0.02)\%
\label{e4pi130}
\end{equation}
and with a temperature of $70$ MeV to be
\begin{equation}
\epsilon_\mathrm{tot}= (0.24\pm0.01)\%\;.
\label{e4pi70}
\end{equation}
Up to now a single-$\phi$ has been used to derive the figures supplied
in eqs.~(\ref{edet}) through~(\ref{e4pi70}), but to really arrive at
the final ones the additional bearing of the whole nuclear event
surrounding the $\phi$-meson has to be included.
The lack in CDC/Barrel matching, due to the double hits in the
latter, can be directly estimated from the data.
From the simulation the expected Barrel detection efficiency for one
charged kaon in the appropriate polar angle acceptance is
$(84\pm2)\%$, owing to the azimuthal coverage and edge effects.
It is difficult to compare safely with the experimental corresponding
one, because $K^+$ and $K^-$ can be cleanly tagged only with $Bmass$
and because of the low statistics.
We opt to refer to a sample of well sorted protons in the CDC, in a
similar momentum range of $0.4-0.6$ GeV/c, although the energy-loss is
not alike.
The measured value is then of $(85\pm 2)\%$ for few-track events,
in good agreement with the prediction.
On the other hand, for multiplicities, typically associated with
$\phi$-candidates, it drops to $(81\pm 2)\%$.
We introduce a correction to $\epsilon_\mathrm{det}$ given by the
square of their ratio
\begin{equation}
\epsilon_\mathrm{mtrk}= (91\pm3)\%\;.
\label{emtrk}
\end{equation}
This is just a lower limit on the reduction in the reconstructed
fraction induced by the presence of all the collision products, it is
nevertheless very well compatible to the embedded-$\phi$ GEANT outcome
of $(92\pm 5)\%$ obtained with a thermal source of $\phi$ at a
temperature of $130$ MeV superimposed on IQMD
\nuc{}{Ni}+\nuc{}{Ni} events~\cite{Mangiarotti:00}.
Such agreement suggests that double hits in the Barrel are the main
origin of failure caused by a realistic multiplicity.
Instead changes in behavior of the CDC tracking code seem to be minor,
confirming its reliability.

The maximum systematic error for the extraction of the efficiency from
the simulation is estimated to be $40\%$ comprehensive of all
contributions: different behavior of each cut in the Monte-Carlo
environment and in the data, small discrepancies in the geometry of
the real and the simulated detector and the influence of the nuclear
event surrounding the $\phi$ (see appendix~\ref{syserr}).

\section{$\phi$-meson production yield}
\label{pmpy}

We are now in a position to deduce the $\phi$-meson yield in central
\nuc{}{Ni}+\nuc{}{Ni} collisions.
Let us focus on the phase space portion relevant for the detector
subsystem (see table~\ref{tacc}) to initially avoid the uncertainty in
the source temperature.
Using the number of $\phi$-candidates of $23$ (see
sec.~\ref{phiid}) obtained from $4.7\cdot10^6$ central events (see
sec.~\ref{evtse}), the $70\%$ detection efficiency (see
eq.~(\ref{edet})) diminished under the realistic track multiplicity
(see eq.~(\ref{emtrk})) and taking into account the partial width of
the $\phi$ into the $K^+\,K^-$-channel of $49.1\%$, we find the
$\phi$-production probability in the CDC/Barrel acceptance per central
event ($0<b<\approx 3.3$ fm)
$P_\mathrm{acc}=(1.6\pm0.5\pm0.8)\cdot10^{-5}$.
The first error is statistical and the second systematic, calculated
with the quadratic and linear sum, respectively, of the contributions
of the data (see sec.~\ref{phiid}) and the simulation (see
sec.~\ref{feff} and appendix~\ref{syserr}).

An extrapolation to a full azimuthal angle coverage of the Barrel and
no edge effects can be performed with $\epsilon^\star$ (see
table~\ref{tacc} and appendix~\ref{filter}): this benefits that it is
independent from the chosen source temperature
\begin{equation}
P_{2\pi}=P_\mathrm{acc}/\epsilon^\star=
(1.9\pm0.6\pm0.95)\cdot10^{-5}\;,
\label{2pi}
\end{equation}
where the errors have an equal meaning as in the previous
$P_\mathrm{acc}$.
Paying attention to the kaon decay in flight (as described in
appendix~\ref{filter}), $P_{2\pi}$ can be immediately compared with
theoretical predictions within the $\vartheta_\mathrm{lab}$ angle and
total momentum acceptances.

If the assumption of a source temperature of $130$ MeV is made (see
sec.~\ref{maxac}), for the yield estimated in the full solid angle, it
is found
\begin{equation}
P_{4\pi\mbox{ @ $130$ MeV}} = P_\mathrm{acc}/
\epsilon_{\mathrm{max @ 130 MeV}}
= (1.2\pm0.4\pm0.6)\cdot10^{-3}\;.
\label{4pi130}
\end{equation}
Finally for a source temperature of $70$ MeV (see sec.~\ref{maxac}),
the same quantity of eq.~(\ref{4pi130}) amounts instead to
\begin{equation}
P_{4\pi\mbox{ @ $70$ MeV}} = P_\mathrm{acc}/
\epsilon_{\mathrm{max @ 70 MeV}}
= (4.5\pm1.4\pm2.2)\cdot10^{-3}\;.
\label{4pi70}
\end{equation}
The errors in eqs.~(\ref{4pi130}) and~(\ref{4pi70}) have been derived
like in eq.~(\ref{2pi}).

The very preliminary result announced earlier in~\cite{Herrmann:96}
is, when compared to eq.~(\ref{4pi130}), smaller by a factor of
$\approx 2.5$: this has to be ascribed to the previous version of the
Monte-Carlo, less accurate than the current one and not yet tuned
through a careful comparison with the real response as in
sec.~\ref{tbiractte} and~\ref{tsotdattsc}.
We underline that the actual size of candidates sample has not been
changed with respect to both~\cite{Herrmann:96}
and~\cite{Mangiarotti:00}.
It is the extracted efficiency that was different, demonstrating the
importance of a sound study about the agreement between simulation and
experiment for the observables necessary to sort out the kaons and
showing how appropriate it is to estimate a maximum possible
systematic errors (see appendix~\ref{syserr}) as attempted in
eq.~(\ref{2pi}) through~(\ref{4pi70}).

The $\phi/K^-$ ratio extrapolated to the full solid angle can now be
given. 
The $K^-$-mesons production probability in $4\pi$ can be deduced to be
$(2.1\pm 0.4)\cdot 10^{-3}$, using the rapidity distribution measured
by KaoS~\cite{Menzel:00} also for \nuc{58}{Ni}+\nuc{58}{Ni} at $1.93$
AGeV and for central collisions, but with the KaoS experiment
centrality selection ($0<b<4.4$ fm) and assuming a Gaussian shape.
Scaling this value with the mean number of participants as
$A_\mathrm{part}^{1.8\pm0.3}$~\cite{Barth:97}, the corresponding one
to the FOPI experiment central trigger is $(2.6\pm 0.5)\cdot
10^{-3}$.
If the last procedure is applied to $K^+$-mesons consistency is found
between the KaoS data~\cite{Menzel:00} and those of
FOPI~\cite{Best:97}.
The $\phi/K^{-}$-ratio is then $(0.44\pm0.16\pm0.22)$ and
$(1.7\pm0.6\pm0.85)$ at $130$ MeV and $70$ MeV temperature,
respectively.
Hence referring to the highest, but still reasonable of the two, and
minding at the branching ratio, it can be established that at least
$\approx 20\%$ of the $K^-$-mesons arise from the decay of an
intermediate $\phi$.

\section{Comparison with theoretical models}

The first detailed theoretical study of $\phi$-meson yield in the
\nuc{}{Ni}+\nuc{}{Ni} system at $1.93$ GeV was that of Chung, Li and
Ko~\cite{Chung:97a,Chung:97b,Chung:98}.
Their relativistic transport model RVUU describes the rapidity
distributions of $p$ and $\pi$ data and the collective flow reasonably
well.
For the $\phi$-meson production, the calculation foresees four
scenarios: i) no in-medium effects, ii) in-medium effects only on the
kaon masses, iii) in-medium effects on the kaon masses and through
this on the $\phi$ decay width and iv) in-medium effects also on the
$\phi$-meson mass.
The parameterization of~\cite{Hatsuda:92} is taken for the in-medium
modification of the $\phi$-meson mass together with the elementary
cross sections from their own model~\cite{Chung:97a}.
Within this frame the dominant channel for $\phi$-production is
$\pi N\rightarrow \phi N$.
The sensitivity was assessed relative to the free scenario i) for the
$K^-$-meson to be a factor of $3$ more in case ii), iii) and iv) and
for the $\phi$ to be a factor of $2$ reduction in case iii) and $10\%$
less in case iv).
Their expectation for the $\phi$-meson occurrence probability in
central collisions is $2.3\cdot10^{-4}$ in case iv).
It is extracted from~\cite{Chung:97b} with the appropriate mean over
$b$ (eq.~(\ref{mpb})).
Assuming a $\phi N$ elastic cross section of $0.56$ mb (their
preferred choice) and of $8.3$ mb (the maximum compatible with the
$\phi$-meson photo-production data) the deduced final $\phi$ kinetic
energy distribution exhibits an inverse slope of $110$ and $130$ MeV,
respectively.
Adopting a temperature of $110$ MeV for extrapolation and comparison
to the experiment, the model underpredicts the data by $6$ times.
Such discrepancy is, however, not too surprising when viewing the
large uncertainties in the parameterization, close to threshold, of
the elementary cross sections.

Recently Barz, Z\'et\'enyi, Wolf and
K\"ampfer~\cite{Zetenyi:02,Barz:02} have been engaged in a new effort,
profiting of the measurements now available for the elementary
$pp\rightarrow\phi pp$ reaction at $\sqrt{s}-\sqrt{s_\mathrm{th}}=83$
MeV~\cite{Balestra:01}.
It is based mainly on their previous work~\cite{Barz:01} about the
role of elementary three body interactions, where the intermediate
particles can be off-shell.
It is found that in current microscopic transport models a small error
is done considering successive two body collisions, where every
particle is on-shell, if all the relevant intermediate steps are
included.
They point at the $B\rho$ and $\pi\rho$-channels as important ones,
neglected in any preceding investigation, and develop their own
predictions starting from this remark.
The in-medium effects are accounted for the $\phi$ alone with the same
result of~\cite{Hatsuda:92} as in the calculations exposed
before~\cite{Chung:97b}.
The $\phi N$ elastic cross section is placed at $0.5$ mb, near to the
preferred estimate of~\cite{Chung:97b}.
The awaited $\phi$-production probability for central collisions
($9\%$ of the total cross section) is $1.7\cdot 10^{-3}$.
The authors apply the FOPI CDC/Barrel geometrical acceptance filter
and obtain a value of $2.7\cdot 10^{-5}$ in reasonable agreement with
the experimental finding of $1.9\cdot 10^{-5}$ (see eq.~(\ref{2pi})).
It is however not clear how well the implementation elaborated for the
moment, can explain the global experimental observables.
In addition it remains to be seen how in-medium modifications of the
kaon properties change the picture, since the $\phi$ yield is coupled
to the sum of the $K^+$ and $K^-$ masses through its decay width.
More work is needed to compare carefully their theory with the
experimental achievement reported here.

A completely different approach is that of the thermal
models~\cite{Cleymans:98a,Cleymans:98b,Cleymans:99} where, assuming a
chemical equilibrium before freeze-out, the abundances can be
predicted (not the yields themselves).
Particular care is to be exercised in the treatment of conservation
laws like baryon number, electric charge and strangeness.
In the thermodynamic limit both canonical and grand canonical
formulations are equivalent, but for small systems, however, the
discrepancies are large.
In fact in general all these quantum numbers should be handled in the
canonical treatment (implying exact conservation) as has been proven
necessary for strangeness, while for baryon number and electric
charge the grand canonical one (implying conservation in mean) is
adequate~\cite{Cleymans:98a}.
If also strangeness is described with the latter, the data in the SIS
energy regime are not matched.
Determining the temperature and baryonic chemical potential from the
$d/p$, $\pi^+/p$ and $K^+/\pi^-$ relative experimental intensities,
the maximum of the $\phi/K^-$ one, that can be accommodated, is of the
order of $0.1$~\cite{Cleymans:99}.
The last value is compatible with the figures of sec.~\ref{pmpy},
within the large systematic and statistical errors, only if a
temperature of at least $130$ MeV is hypothesized.
This situation illustrates the limitations of the actual small
acceptance measurement where the $\phi$ source is not well
characterized and of this class of models where the species kinetic
temperatures are poorly predictable.

\section{Conclusions}

A first experimental production probability of sub-threshold
$\phi$-meson in heavy ion collisions has been obtained.
Before solely $K^+$ and $K^-$ data were available.
With a pseudo-vector meson a new sector of the theoretical
machinery, proposed to describe in-medium effects, has been
tested.
It supports the emerging awareness of a strong interconnection between
the $K^+$, $K^-$ and $\phi$-channels.

Here only the data gained from the CDC/Barrel subsystem are discussed;
the effort necessary to make a quantitative statement for the
probability of occurrence in central collisions is described.
The creation of a $\phi$-meson is a very rare event at sub-threshold
energies: out of $4.7\cdot10^6$ events in the sample of central
collisions, barely $23$ candidates were found after background
subtraction.
Despite the poor statistics, a signal to background ratio of $\approx
1$ could be achieved when applying appropriate cuts.

An important part of the effort has been the estimation of the
efficiency that was splitted into two parts: one,
$\epsilon_\mathrm{max}$, due to the geometrical acceptance, allowed
momentum range and $K^+$ and $K^-$ decay in flight and a
second, $\epsilon_\mathrm{det}$, originating in the identification
requirements necessary to decrease the background.
While $\epsilon_\mathrm{max}$ is between $1.35\%$ and $0.36\%$
(depending on the source temperature), showing that the greatest loss
in $\phi$ events is to be attributed to the localization of the phase
space covered by the detector around target rapidities,
$\epsilon_\mathrm{det}$ is $70\%$ indicating the validity of the cuts
chosen in reducing the background without rejecting too many good
events.

Furthermore $\epsilon_\mathrm{det}$ was proven to be insensitive,
within the statistical error of the simulations, to the source
temperature (and it could be argued even from the shape of a
moderately anisotropic one); on the contrary
$\epsilon_\mathrm{max}$ varies by a factor of $4$ when the temperature
changes from $70$ to $130$ MeV.
The comparison with calculations or new results is hence greatly
simplified: in fact while $\epsilon_\mathrm{max}$ can easily be
evaluated along the guidelines of appendix~\ref{filter} by other
groups, $\epsilon_\mathrm{det}$ can exclusively be extracted from a
complete simulation of the detector response, requiring a specific
knowledge of the apparatus; it has been deduced once for all in the
present work.

Future confrontations will benefit from a thorough discussion of the
possible sources of systematic error: a value of $40\%$ and $9\%$ has
been given for simulation and data, respectively.

This measurement suggests that, near threshold, a large fraction of
the $K^-$-mesons (at least $20\%$) arise from the decay of an
intermediate $\phi$-meson.
As a consequence it would be important to verify, both experimentally
and theoretically, to what extent the $K^-$ phase space distribution
and the $\phi/K^-$-ratio can be contaminated from the
$\phi$-production channel.
The discrepancies with the predictions based on microscopic transport
codes can partially be ascribed to insufficient knowledge of the
elementary cross sections.
A possible role of intermediate $\rho$-mesons production for the $\phi$
yield appears to be an interesting aspect of investigation.
The $\phi/K^-$-ratio seems to be hardly compatible with thermal
models.

In order to improve the knowledge on the experimental side, a more
complete coverage of the acceptance is needed.
Especially, the access to the region of low transverse momenta in the
proximity of mid-rapidity is important for a reliable reproduction of
the total yield of such rare particles produced in central heavy-ion
collisions.
A first step along this way is taken by the ongoing effort to include
the Helitron/Forward Wall data~\cite{Kotte:00}.
In a near future the FOPI collaborations plans to accumulate more
data with an increased statistics on $\phi$-production, when the
undertaken upgrade program will be finished.
On the other hand, the HADES collaboration will be able to detect the
$\phi$-meson through the leptonic channel in search for direct
evidences of in-medium effects.
Both plans make the subject open for promising new developments.

\section*{Acknowledgments}

One of the authors (A.M.) is in debt with M.~Bini, G.~Casini, A.~Olmi,
G.~Pasquali, A.~Perego, S.~Piantelli, G.~Poggi, P.~Sona and
A.~Stefanini for helpful discussions and assistance.
He is also grateful to his family, V.~Biagioli, L.~Cavallini,
L.~Mannucci, F.~Masi, C.~Mastella, E.~Pagliai, P.~Panicucci and
A.~Seracini for encouragement.

This work has been supported under the EC contract HPRI-CT-1999-00001
and in part by the German BMBF under contracts 06HD953,
RUM-005-95/ RUM-99/010, POL-119-95, UNG-021-96 and RUS-676-98 and by
the Deutsche Forschungsgemeinschaft (DFG) under projects
436-RUM-113/10/0, 436-RUS-113/143/2 and 446-KOR-113/76/0.
Support has also been received from the Polish State Committee of
Scientific Research, KBN, from the Hungarian OTKA under grant T029379,
from the Korea Science and Engineering Foundation under grant
20015-111-01-2, from the agreement between GSI and CEA/IN2P3 and from
the PROCOPE Program of DAAD.


\newpage
\begin{table}[H]
\begin{center}
\begin{tabular}{||c|rl||}\hline\hline
$B$               & .6   & Tesla \\\hline\hline
\multicolumn{3}{||c||}{CDC}\\\hline\hline
$R_\mathrm{max}$         & $20.7$       & cm\\
$R_\mathrm{min}$         & $80.1$       & cm\\
$\vartheta_\mathrm{min}$ & $32.7$       & $^\circ$\\
$\vartheta_\mathrm{max}$ & $130-154$    & $^\circ$\\\hline\hline
\multicolumn{3}{||c||}{Barrel}\\\hline\hline
$R_\mathrm{Bar}$         & $111.5$      & cm\\
$\vartheta_\mathrm{min}$ & $39.2$       & $^\circ$\\
$\vartheta_\mathrm{max}$ & $132.9$      & $^\circ$\\
$z_\mathrm{min}$         & $-103.5$     & cm\\
$z_\mathrm{max}$         & $136.5$      & cm\\
$\epsilon^\star$         & $81\pm 2.5$  & $\%$\\\hline\hline
\multicolumn{3}{||c||}{Momentum}\\\hline\hline
$p_\mathrm{max}$         & $0.6$        & GeV/c\\\hline\hline
\end{tabular}
\end{center}
\vspace{0.8cm}
\caption{Geometrical acceptance of the CDC and the Barrel.}
\label{tacc}
\end{table}

\begin{table}[H]
\begin{center}
\begin{tabular}{||c|rcl|rcl||}\hline\hline
\multicolumn{7}{||c||}{Acceptance}\\
\multicolumn{1}{||c}{}& \multicolumn{3}{c}{Data} &\multicolumn{3}{c||}{GEANT}\\\hline\hline
$p$             & $<$   & $0.6$  & GeV/c    & $<$   & $0.6$  & GeV/c\\\hline\hline
\multicolumn{7}{||c||}{Tracker}\\
\multicolumn{1}{||c}{}& \multicolumn{3}{c}{Data} &\multicolumn{3}{c||}{GEANT}\\\hline\hline
$\Delta_{xy\,hit}$ & $\pm$ & $0.11$ & cm       & $\pm$ & $0.11$ & cm\\
$\Delta_{z\,hit}$  & $\pm$ & $35$   & cm       & $\pm$ & $35$   & cm\\
$\Delta_{z}$       & $\pm$ & $25$   & cm       & $\pm$ & $19$   & cm\\
$\Delta_{\phi}$    & $\pm$ & $2$    & $^\circ$ & $\pm$ & $2$    & $^\circ$\\\hline\hline
\multicolumn{7}{||c||}{Analysis}\\
 \multicolumn{1}{||c}{}&\multicolumn{3}{c}{Data}&\multicolumn{3}{c||}{GEANT}\\\hline\hline
$n_{hit}$       & $>$     & $30$  &     & $>$   & $30$  &   \\
$d_0$           & $\pm$   & $0.5$ & cm  & $\pm$ & $0.5$ & cm\\
$z_0$           & $\pm$   & $10$  & cm  & $\pm$ & $20$  & cm\\
$Cmass$         & \multicolumn{3}{|c|}{$m>0.37$ and $m<0.8$ GeV} & \multicolumn{3}{|c||}{$m>0.37$ and $m<0.8$ GeV}\\
$Bmass$         & \multicolumn{3}{|c|}{$|m-0.494|<0.1$ GeV} & \multicolumn{3}{|c||}{$|m-0.494|<0.1$ GeV}\\\hline\hline
\end{tabular}
\end{center}
\vspace{0.8cm}
\caption{Momentum, tracker and identification cuts used for the
analysis of the data and of the simulation.}
\label{tcut}
\end{table}

\begin{landscape}
\begin{table}[H]
\begin{center}
\begin{tabular}{||r||r|r|r|r|r|r|r||}\hline\hline
  T  & \multicolumn{1}{c}{$\epsilon_\mathrm{CDC}$ $(\%)$} & \multicolumn{1}{|c}{$\epsilon_\mathrm{Bar}$ $(\%)$} & \multicolumn{1}{|c}{$\epsilon_\mathrm{max}$ $(\%)$} & \multicolumn{1}{|c}{$\epsilon_\mathrm{Bar}$ $(\%)$} & \multicolumn{1}{|c}{$\epsilon_\mathrm{max}$ $(\%)$} & \multicolumn{1}{|c}{$\epsilon_\mathrm{tot}$ $(\%)$} & \multicolumn{1}{|c||}{$\epsilon_\mathrm{det}$ $(\%)$}\\
 & \multicolumn{1}{|c}{No decay} & \multicolumn{1}{|c}{No decay} & \multicolumn{1}{|c}{No decay} & \multicolumn{1}{|c}{Decay} & \multicolumn{1}{|c|}{Decay} & \multicolumn{1}{|c|}{Decay} & \\\hline\hline
130  & $17.03\pm 0.09$ & $7.20\pm 0.06$   &  $3.78\pm 0.04$  & $3.26\pm 0.04$   & $1.35\pm 0.02$   & $.950\pm 0.02$     & $70\pm 2$\\
110  & $13.19\pm 0.07$ & $5.17\pm 0.04$   &  $2.93\pm 0.03$  & $2.26\pm 0.03$   & $1.07\pm 0.02$   & $.751\pm .016$     & $70\pm 2$\\
 90  & $ 9.36\pm 0.05$ & $3.20\pm 0.03$   &  $2.02\pm 0.02$  & $1.31\pm 0.02$   & $0.67\pm 0.01$   & $.475\pm .012$     & $71\pm 2$\\
 70  & $ 5.53\pm 0.04$ & $1.63\pm 0.02$   &  $1.10\pm 0.02$  & $0.64\pm 0.01$   & $0.36\pm 0.01$   & $.242\pm .007$     & $67\pm 3$\\\hline\hline
\end{tabular}
\end{center}
\vspace{0.8cm}
\caption{Geometrical $\epsilon_\mathrm{CDC}$ and
$\epsilon_\mathrm{Bar}$, maximum $\epsilon_\mathrm{max}$, total
$\epsilon_\mathrm{tot}$ and detection $\epsilon_\mathrm{det}$ (see
eq.~(\ref{eff})) efficiencies for four thermal source temperatures.
The geometrical contribution is given for the CDC alone
($\epsilon_\mathrm{CDC}$) and the CDC/Barrel combination
($\epsilon_\mathrm{Bar}$).
The latter $\epsilon_\mathrm{Bar}$ includes the not complete azimuthal
coverage of the Barrel and edge effects (see $\epsilon^\star$ of
table~\ref{tacc} and appendix~\ref{filter}).
If, within the acceptance of the CDC/Barrel subsystem
($\epsilon_\mathrm{Bar}$), the upper momentum limit at $0.6$ GeV/c is
imposed, $\epsilon_\mathrm{max}$ is obtained.
For $\epsilon_\mathrm{Bar}$ and $\epsilon_\mathrm{max}$ the reduction
due to the kaon decay in flight during their path to the Barrel is
considered.
All figures assume that the only open channel for the $\phi$-meson is
the $K^+\,K^-$ one.}
\label{teff}
\end{table}
\end{landscape}

\newpage

\begin{figure}[H]
\centering\mbox{\epsfig{file=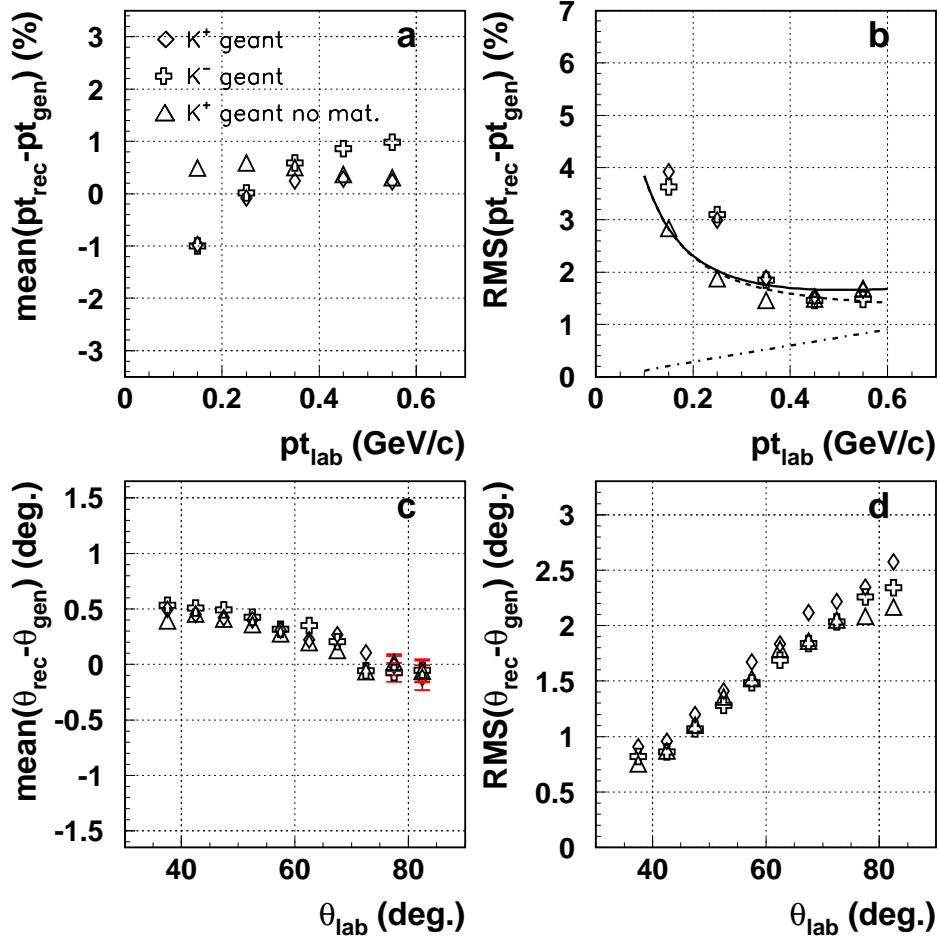,width=1.\textwidth}}
\caption{Transverse momentum $p_t$ and longitudinal angle
$\vartheta_\mathrm{lab}$ systematic shift and resolution from the
Monte-Carlo.
In addition to $K^+$ and $K^-$-mesons from the standard environment
used in the present work, also $K^+$ from a simulation
without target and chamber entrance materials are considered for
comparison.
In panel~b the $p_t$ root mean square deviation is compared with what
is expected from Gluckstern's formulas~\cite{Gluckstern:63} describing
the effect of multiple scattering in the CDC gas (dashed line) and of
a transverse hit position uncertainty of $400$ $\mu$m (see
fig.~\ref{gcdc}~a) under the vertex constraint (dot-dashed line).
The two contributions are added quadratically (continuous line).}
\label{gptthe}
\end{figure}

\newpage
\appendix
\section{Direct calculation of the geometrical acceptance}
\label{filter}

If the $\phi$-meson source is not spherical or the temperature is
outside the range of table~\ref{teff} (so that an interpolation can
can not be used) $\epsilon_\mathrm{max}$ has to be calculated
from the events generated by the assumed model.
We briefly describe the filter procedure, with the related setup
parameters, which is needed to compare future data or the prediction
of a microscopic transport code with the $\phi$-production probability
measured here in the CDC/Barrel acceptance.
After the $\phi$ has decayed into a $K^+$ and a $K^-$-meson when both
have a momentum smaller than $0.6$ GeV/c, for each of them it is
necessary to determine if it hits the nominal Barrel surface.
The latter is defined as a cylinder located at a radius
$R_\mathrm{Bar}$ with a $z$ extension from $z_\mathrm{min}$ to
$z_\mathrm{max}$.
First for the inverse of the radius of curvature in the transverse
plane $k$, that can be obtained from eq.~(\ref{pt}), the condition
$2/k>R_\mathrm{Bar}$ has to be fulfilled.
In this case the length $l_t$ of the transverse projection of the
track from the origin to the Barrel can be calculated as
\begin{equation}
l_t=\frac{2}{k} \arcsin\left(\frac{R_\mathrm{Bar}\,k}{2}\right)
\end{equation}
and the $z$-coordinate of the hit in the scintillator bar is
\begin{equation}
z= h\,l_t\;,
\label{z}
\end{equation}
where $h$ is the helix pitch defined by $h=p_z/p_t$.
Second the value given by~(\ref{z}) has to lie inside the interval
$z_\mathrm{min}\leq z\leq z_\mathrm{max}$
with the $z_\mathrm{min}$ and $z_\mathrm{max}$ of table~\ref{tacc}.
To account for the possible decay in flight, each kaon must be
weighted with a survival probability
\begin{equation}
\exp\left(-\frac{l\,m_K}{p_K\,\,c\,\tau_K}\right)\;,
\label{pdec}
\end{equation}
where $m_K$, $\tau_K$, $p_K$ are the kaon mass, mean life and momentum,
respectively and $l$ is the corresponding distance of flight to the
Barrel
\begin{equation}
l=l_t\,\sqrt{1+h^2}\;.
\end{equation}
After the weighting with the product of the quantities~(\ref{pdec})
for each particle of the couple, the final outcome delivers
$\epsilon_\mathrm{max}^\star$, where
\begin{equation}
\epsilon_\mathrm{max}=\epsilon_\mathrm{max}^{\star}\cdot\epsilon^\star\;.
\end{equation}
In order to obtain $\epsilon_\mathrm{max}$ of eq.~(\ref{eff}) and
table~\ref{teff}, a reduction factor $\epsilon^\star$, comprehensive
of both kaons, has to be applied to compensate for the incomplete
azimuthal coverage of the Barrel due to the regions occupied by the
CDC support structures and to the inter-module and inter-bar
spacings.
Because the depth of the scintillator bar is bigger than the width,
edge effects are important at the level of a few percent
(see sec.~\ref{tsotdattsc}) and have to be included as well.
The value of $\epsilon^\star$ has been estimated through the ratio
between the output of the filter just outlined and of GEANT where the
geometry is defined accurately implementing separately each
scintillator bar with its appropriate positions and dimensions.
It turns out from the simulation that
$\epsilon^\star$ is independent from the source temperature in the
considered range, within one standard deviation, and can be determined
once for all as $(81\pm2.5)\%$ (see table~\ref{tacc}).
In a comparison of theory to the experiment one can benefit of
eq.~(\ref{2pi}) and barely supply $\epsilon_\mathrm{max}^\star$ so
that
\begin{equation}
P_{4\pi}=P_{2\pi}/\epsilon_\mathrm{max}^\star\;.
\end{equation}

\section{Estimate of the systematic errors}
\label{syserr}

Our estimate of the efficiency is biased by systematic errors of
distinct types: a non-identical behavior of every cut in the data and
simulation, small geometrical discrepancies between the real detector
and the simulated one, the influence of the nuclear event surrounding
the $\phi$-meson.
We discuss each of them ($d_0$, $z_0$, $Bmass$, $Cmass$, $\Delta_z$,
$\Delta_\varphi$, $n_{hits}$, upper momentum, Barrel geometry,
multi-track environment) in turn.

Beginning with the $K^+\,K^-$-meson selection requirements, the ones
on $d_0$ and $z_0$ are assumed to contribute negligible systematic
distortions.
As a matter of fact, the distribution of the first is well reproduced
and the bump starting at $-0.5$ cm in the simulation is due to mirror
tracks that have to be discarded (see sec.~\ref{tsotdattsc}).
Any eventual correction arising from the improper purge of such cases
is assumed minor.
The latter does not reject good events and is wide enough to
accommodate the small discordances in the experimental and awaited
shapes, once the condition imposed on $z_0$ in the analysis of GEANT
is released as given in table~\ref{tcut}.

The $Bmass$, $Cmass$ and $\Delta_z$ observables obey a Normal law with
good approximation in the data and in the simulation, besides such
assumption is legitimate for the estimation of systematic errors.
As long as the real and calculated spectra have a Gaussian shape there
could be only two sources of concern about the effect of a fixed
accepted window: a difference in the mean or in the standard deviation
$\sigma$.
In truth if both are not quite similar the integral inside the
same range is not equal, as can be easily obtained from the error
function.
The discrepancy in the efficiency of a cut between the simulation and
the data translates directly by twice the amount (once for each kaon
of the pair) to the correctly identified fraction
$\epsilon_\mathrm{det}$.
The change for variations in the mean or the $\sigma$ judged
reasonable from our previous discussion (see sec.~\ref{tsotdattsc}) is
written up in the rightmost column of table~\ref{tsys} in percentage
of the number of events allowed by the selection.
Since the domain of uncertainty quoted in mean or $\sigma$ are the
maximum expected ones, an analogous statement is true also for the
systematic error (in the same spirit for evaluating the effect of the
shift in mean always the larger of the two cited $\sigma$'s is
preferred).
In the case of $Cmass$ the permitted region is widely asymmetric and
solely the lower exclusion value matters, this peculiarity was
accounted.

For $\Delta_z$, the width of the required interval used in the
analysis of the data was rescaled for the simulation one.
A small residual repercussion of the systematic discrepancy in the
standard deviation is expected as compared to the others where the
window is left untouched, here we will neglect it.
Even if $\Delta_\varphi$, partially reflecting the geometrical shadow
of the scintillator bar, does not properly follow a Normal law,
a Gaussian is not a too bad approximation for estimating the
systematic error.
It turns out to be very small, after all the cut is sufficiently wide,
and it can be neglected (see table~\ref{tsys}).
The two rms of $\Delta_\varphi$ are well compatible with each other
and no adjoined systematic distortions are considered from this side.
The alteration in rejection caused by the dissimilarities in the shape
of real and foreseen $\Delta_z$ and $\Delta_\phi$ distributions is in
general not expected to be the major contribution to the systematic
error in the matching ratio.
In fact the corresponding accepted windows are $\approx 3\,\sigma$
wide and, as already discussed (see sec.~\ref{tsotdattsc}), the origin
of the loss is rather the scintillator efficiency coupled to the edge
effects for curved tracks.
However, if the matching fraction for single kaons predicted by the
GEANT model $(84\pm 2)\%$ and the measured one for well defined CDC
protons tracks $(85\pm 2)\%$ are compared (see sec.~\ref{feff}), the
discordance of $1\%$ is not statistically meaningful.
So no further contribution is introduced from the systematic
limitations in the implementation of all the detailed processes
affecting $\Delta_z$ and $\Delta_\phi$. 

The invariant mass spectrum, being dominated by the CDC momentum
resolution, is also approximately Gaussian (see sec.~\ref{feff}).
No statistically significant disagreement exists between measurements
and Monte-Carlo either for the mean or for the width, so the
associated systematic error is assumed to be negligible.

The probability for a track to have an assigned number of hits does of
course not follow a Normal law but is rather asymmetric with a tail
towards low hit multiplicities (even for kaons reaching the Barrel)
populated by fake tracks alone in the simulation and also by pion
background in the experiment.
It is difficult to quantify the reliability of the description of this
quantity because it depends on many subtle particularities of the
detector response, so an estimate (see table~\ref{tsys}) equal to the
remainder of what is rejected in the simulation ($3\%$) and in the
data ($5.5\%$) was taken, it can not be exceed by the true
discrepancy.

We interpret systematic errors as reasonable estimates of the possible
displacement of the true result by effects not completely controllable
and opt for a linear sum with the aim of arriving at trustable gap in
which the final efficiency should lie.
If the disagreement with a calculation is outside this, serious doubts
should be shed on the theory itself.
Adding all the discussed sources a maximum systematic error of $13\%$
is found for one kaon and of $26\%$ for the pair, i.e. for
$\epsilon_\mathrm{det}$.

The margin of uncertainty due to the upper momentum limitation was
estimated varying the cut of $\pm1\%$, that is roughly the systematic
distortion of the momentum reconstruction (see fig.~\ref{gptthe}) and
taking the biggest change (see table~\ref{tsys}) in the size of the
selected $\phi$-sample (the source temperature was $130$ MeV).
Including this last contribution gives an expectation of $28\%$
systematic error imputable to imperfect characterization of the kaon
selection constraints behavior.

The difficulty in the determination of the geometrical acceptance can
be reduced to the problem of knowing with enough precision the
respective positions of the target and the Barrel, whose polar angle
coverage in the forward direction is smaller than that of the CDC.
Since the target is replaced after a set of runs, its displacement has
to be deduced each time from relative measurements of the parts of its
holder system, a maximum safety boundary of $\pm1$ cm is appropriate.
A simulation was run with a thermal source of $\phi$-mesons at $130$
MeV temperature and the Barrel was moved of $\pm1$ cm, the influence
on $\epsilon_\mathrm{max}$ is reported in table~\ref{tsys}.

For estimating the possible systematic mismatch in the reduction of
detection efficiency by the multi-track environment, the difference in
the total percentage of correctly identified events in the
single-$\phi$ and embedded-$\phi$ Monte-Carlo is indicated in the
rightmost raw of table~\ref{tsys}.

A maximum comprehensive systematic error for the extraction of the
efficiency from the simulation of $40\%$ is estimated.


\begin{landscape}
\begin{table}[H]
\begin{center}
\begin{tabular}{||c|rcl|rcl|rcl|rl||}\hline\hline
    & \multicolumn{3}{c|}{cut} & \multicolumn{3}{c|}{data} & \multicolumn{3}{c|}{GEANT} & \multicolumn{2}{c||}{syst. error}\\
$Bmass$          & $|Bmass/m_K-1|<$    & $20$  & $\%$   & $\langle m_b \rangle$      & $\approx$ & $0\%$  & $\langle m_b \rangle$      & $=$ & $-2\%$ & $0.45$ & $\%$\\
                 &                     &       &        & $\sigma(m_b)$              & $\approx$ & $8\%$  & $\sigma(m_b)$              & $=$ & $10\%$ & $3.3$  & $\%$\\
$Cmass$          & $Cmass/m_K-1>$      & $-25$ & $\%$   & $\langle m_c \rangle$      & $\approx$ & $5\%$  & $\langle m_c \rangle$      & $=$ & $10\%$ & $3.6$  & $\%$\\
                 &                     &       &        & $\sigma(m_c)$              & $\approx$ & $8\%$  & $\sigma(m_c)$              & $=$ & $13\%$ & $2.6$  & $\%$\\
$\Delta_z$       & $|\Delta_z|<$       & $25$  & cm     & $\langle \Delta_z \rangle$ & $\approx$ & $4$ cm & $\langle \Delta_z \rangle$ & $=$ & $0$ cm & $0.27$ & $\%$\\
$\Delta_\varphi$ & $|\Delta_\varphi|<$ & $2$ & $^\circ$ & $\langle \Delta_\varphi \rangle$ & $\approx$ & $0^\circ$ & $\langle \Delta_\varphi \rangle$ & $\pm$  & $.2^\circ$ & $0.06$ & $\%$\\\hline
hit mul.         & $n_{hit}>$          & $30$  &        & \multicolumn{3}{|c|}{rejection $5.5\%$}         & \multicolumn{3}{|c|}{rejection $3\%$}     & $2.5$  & $\%$\\\hline
$p$              & $p<$                & $0.6$ & GeV/c  & \multicolumn{6}{c|}{variation of $\pm 1\%$ @ $T=130$ MeV}                                   & $2.6$  & $\%$\\\hline
\multicolumn{4}{||c}{Barrel geometry}  & \multicolumn{6}{|c|}{ Barrel displacement of $\pm 1$ cm @ $T=130$ MeV}                                       & $3.8$  & $\%$\\
\multicolumn{4}{||c}{multi-track env.} & \multicolumn{6}{|c|}{embedded $\phi$ simulation @ $T=130$ MeV}                                           & $8$    & $\%$\\\hline\hline
\end{tabular}
\end{center}
\vspace{0.8cm}
\caption{Contributions to the systematic error.
The values of the selection windows considered are those of the
analysis of the data and the simulation (see table~\ref{tcut}).
For the first four cuts the mean and standard deviation $\sigma$
quoted for the data are based on estimations of the maximum
discrepancy and for the simulation are simply what is obtained at
the highest permitted momentum of $0.6$ GeV/c.
From the difference between the measurements and the Monte-Carlo the
systematic error on the cut effect is calculated assuming a Normal
distribution and given as a percentage of the events allowed.
For the last four contributions the estimation is adapted for each as
explained in the text (see appendix~\ref{syserr}).}
\label{tsys}
\end{table}
\end{landscape}


\newpage
\bibliography{rhicoll,fopidet}
\bibliographystyle{elseart-num}

\end{document}